# Relaxation of two coupled quantum oscillators to quasi-equilibrium states based on path integrals

Illarion Dorofeyev

Institute for Physics of Microstructures, Russian Academy of Sciences,

603950, GSP-105 Nizhny Novgorod, Russia

## Abstract

The paper addresses the problem of relaxation of open quantum systems. Using the path integral methods we found an analytical expression for time-dependent density matrix of two coupled quantum oscillators interacting with different baths of oscillators. The expression for density matrix was found in the linear regime with respect to the coupling constant between selected oscillators. Time-dependent spatial variances and covariance were investigated analytically and numerically. It was shown that asymptotic variances in the long-time limit are always in accordance with the fluctuation dissipation theorem despite on their initial values. In the weak coupling approach there is good reason to believe that subsystems asymptotically in equilibrium at their own temperatures even despite of the arbitrary difference in temperatures within the whole system.

PACS Nos: 05.40.Jc; 42.50.Lc; 05.10.Gg;

## I. Introduction

A quantum oscillator or set of oscillators coupled to heat baths is a traditional problem of open quantum systems. Brownian dynamics of particles is often modeled by coupling the system to a large number of harmonic oscillators. The hamiltonian system composed by a selected particle plus a thermal reservoir followed by reduction with respect to reservoir's variables allow investigating the origin of irreversibility in the dynamics of a quantum system interacting with a heat bat; see for instance [1-10]. In this connection, thermodynamic characteristics of a quantum oscillator coupled to a heat bath, various definitions of the characteristics, some subtleties due to possible approximations were studied in [11-15]. In parallel, another set of adjoint studies were performed from the point of view of the relaxation of open systems to the steady state and description of the nonequilibrium transport phenomena. A problem for coupled oscillators interacting with heat baths in different approximations was considered in [16-21] in order to clarify the approach to equilibrium or to some transient stationary state of the system. It was shown that an arbitrary initial state of a harmonic oscillator state decayes towards a uniquely determined stationary state. An explicit expression for the time evolution of the density matrix when the system starts in a particular kind of pure state was derived and investigated in [22] for the case of a harmonic oscillator with arbitrary damping and at arbitrary temperature. The spatial dispersion in the infinite time limit was shown to agree with the fluctuation-disspation theorem. A study of nonequilibrium properties of a harmonic chain to whose ends independent heat baths are attached was performed in [23]. It was found that as a time goes to infinity, the chain approaches a stationary state regime. An analyze of the nonequilibrium steady states of a one-dimensional harmonic chain of atoms with alternating masses connected to heat reservoirs at unequal temperatures and a heat transfer across an arbitrary classical harmonic network connected to two heat baths at different temperatures were done in [24,25]. An entanglement evolution of two harmonic oscillators under the influence of non-Markovian thermal environments including common or separate baths was studied in [26]. It was found that the dynamics of the quantum entanglement is sensitive to the initial states, the oscillator-oscillator interaction, the oscillator-environment interaction and the coupling to a common bath or to different, independent baths. The non-Markovian dynamics of two independent qubits, each locally interacting with a zero-temperature reservoir was investigated in [27]. The authors found that the non-Markovian effects influence the entanglement dynamics and may give rise to a revival of entanglement even after complete disentanglement for finite time periods. A complete analysis of the evolution of entanglement and quantum discord between two resonant oscillators coupled to a common environment was provided [28-30]. Different phases of evolution including sudden death and revival of entanglement for different models of environments and for different models for the interaction between the system and reservoirs were described. The evolution of quantum states of networks of quantum oscillators coupled with arbitrary external environments was analyzed in [31]. The emergence of thermodynamical laws in the long time regime and some constrains on the low frequency behavior of the environmental spectral densities were demonstrated.

The above mentioned model problems are similar to the natural problems of interaction between systems within different thermostats kept with own temperatures. In series of works it was established that the difference in temperatures yields in new effects within natural systems out of equilibrium. For instance, general expressions for the van der Waals interaction between two molecules having different temperatures were derived in [32-34] and demonstrated that the dispersion potential becomes repulsive when the temperature difference is large enough. A change of the adsorption potential between a molecule and a nonequilibrium semiconductor was found in [35]. Dispersion forces between two molecules that are in relative motion were calculated in [36]. It was found that the nonequilibrium forces may be attractive or repulsive and may be nonconservative. Corresponding problem in physics of the dispersion forces between macroscopic bodies out of thermal equilibrium was investigated in [37-43]. A new distance dependence between a molecule and solid body due to diffference in temperatures was discovered in [39-41, 43]. Analytical expression for mean energy of interaction of two coupled oscillators within independent heat reservoirs of harmonic oscillators in a steady state regime was derived in [44]. It should be emphasized that all above mentioned problems are combined by introducing one essential common feature concerned with stationarity of the systems under study. Obviously, that the problems of the temporal dynamics of interacting systems from some initial time up to some arbitrary time of steady state and the stationarity itself of the systems out of equilibrium are very important points and allow understanding of complex natural phenomena.

This paper addresses to derive an analytical expression for a density matrix of two quantum coupling oscillators within independent heat baths of quantum oscillators. The main goal in this paper is twofold: firstly, to represent analytically the reduced density matrix explicitly extracting the Feynman-Vernon influence functional and suitable propagator for our case, and secondly to demonstrate an existence of the quasi-equilibrium steady state of the system under study.

The paper is organized as follows. In Sec.II we provide a theoretical description based on path integrals to calculate the reduced density matrix of two interacting quantum oscillators in different reservoirs of harmonic oscillators. The Feynman-Vernon influence functional, relative propagator and analytical formulas for the density matrix in the linear regime with respect to the coupling constant and different temporal asymptotes are given in Sec.III. Our conclusions are given in Sec.IV.

## II. Problem statement and solution

We consider two bilinear coupling oscillators, and each of these oscillators, in its turn, is bilinear coupled with separate reservoirs of oscillators. Corresponding Hamiltonian is written as follows

$$\begin{aligned} H &= p_1^2/2m_1 + M_1\omega_{01}^2 x_1^2/2 + p_2^2/2m_2 + M_2\omega_{02}^2 x_2^2/2 - \lambda x_1 x_2 + \\ &+ \sum_{j=1}^{N_1}\left[p_j^2/2m_j + m_j\omega_j^2(q_j - x_1)^2/2\right] + \sum_{k=1}^{N_2}\left[p_k^2/2m_k + m_k\omega_k^2(q_k - x_2)^2/2\right], \end{aligned} \tag{1}$$

where $x_{1,2}\,, p_{1,2}\,, M_{1,2}\,, \omega_{01,02}$ are the coordinates, momenta and masses of the selected oscillators, $\lambda$ is the coupling constant, $q_j\,, p_j\,, \omega_j\,, m_j$ and $q_k\,, p_k\,, \omega_k\,, m_k$ are the coordinates, momenta, eigenfrequencies and masses of bath's oscillators. Further we use the vectors $\vec{R}_1 = \{q_j\} = \{q_1,...,q_{N_1}\}$ and $\vec{R}_2 = \{q_k\} = \{q_1,...,q_{N_2}\}$ for brevity.

Similar Hamiltonians, the nature of coupling, frequencies renormalization, unitary transformations and other points were discussed in detail, for example in [6, 10, 22].

Figure 1 presents a simplified sketch of the problem under consideration. In time $t < 0$ we have uncoupled system of oscillators. Then, the interactions are switched on in the time $t = 0$ and maintained during arbitrary time interval up to infinity. The problem is to find the time-dependent density matrix of two selected coupled oscillators in any moment of time $t \ge 0$.

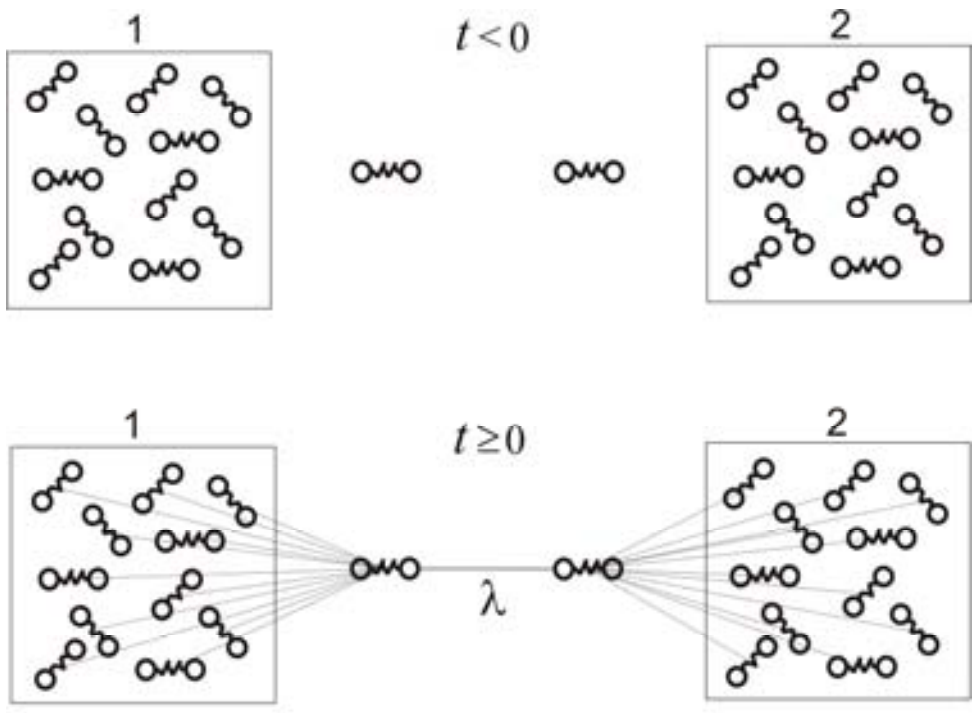

Fig.1

It is well known that the evolution of the total Hamiltonian system "selected two interacting oscillators plus two reservoirs" is described by the equation for the density matrix $W(t)$ of the total system

$$W(t) = \exp(-iHt/\hbar)W(0)\exp(iHt/\hbar), \tag{2}$$

where $W(0)$ is the initially prepared density matrix of the total system.

Using the completeness properties of the eigenfunctions in the coordinate representation the Eq. (2) can be rewritten [45] as follows

$$\begin{aligned}&\left\langle x_1 x_2 \vec{R}_1 \vec{R}_2 \left| W(t) \right| y_1 y_2 \vec{Q}_1 \vec{Q}_2 \right\rangle = \int dx_1' dx_2' dy_1' dy_2' d\vec{R}_1' d\vec{R}_2' d\vec{Q}_1' d\vec{Q}_2' \\ &\left\langle x_1 x_2 \vec{R}_1 \vec{R}_2 \left| \exp(-iHt/\hbar) \right| x_1' x_2' \vec{R}_1' \vec{R}_2' \right\rangle \left\langle x_1' x_2' \vec{R}_1' \vec{R}_2' \left| W(0) \right| y_1' y_2' \vec{Q}_1' \vec{Q}' \right\rangle \left\langle y_1' y_2' \vec{Q}_1' \vec{Q}_2' \left| \exp(iHt/\hbar) \right| y_1 y_2 \vec{Q}_1 \vec{Q}_2 \right\rangle ,\end{aligned} \tag{3}$$

with the limits of integration extended from minus to plus infinity.
The transition amplitudes in Eq.(3) are expressed via the path integrals [46-49, 10]

$$\left\langle x_1 x_2 \vec{R}_1 \vec{R}_2 \left| \exp(-iHt/\hbar) \right| x_1' x_2' \vec{R}_1' \vec{R}_2' \right\rangle \equiv K(x_1, x_2, \vec{R}_1, \vec{R}_2, t; x_1', x_2', \vec{R}_1', \vec{R}_2', 0) = \int \mathcal{D}\bar{x}_1 \mathcal{D}\bar{x}_2 \mathcal{D}\underline{\vec{R}}_1 \mathcal{D}\underline{\vec{R}}_2 \exp\{(i/\hbar) S[\bar{x}_1, \bar{x}_2, \underline{\vec{R}}_1, \underline{\vec{R}}_2]\} \quad , \tag{4}$$

where we designated the integration variables by straight lines beneath the letters for bath's coordinates $\underline{\vec{R}}_{1,2}$ and over the letters for oscillator's coordinates $\bar{x}_{1,2}$ . The integration along all paths is carried out from $\bar{x}_1(0) = x_1'$ to $\bar{x}_1(t) = x_1$, from $\bar{x}_2(0) = x_2'$ to $\bar{x}_2(t) = x_2$, and from $\underline{\vec{R}}_1(0) = R_1'$ to $\underline{\vec{R}}_1(t) = R_1$, from $\underline{\vec{R}}_2(0) = R_2'$ to $\underline{\vec{R}}_2(t) = R_2$.

The backward amplitude in Eq. (3) is

$$\left\langle y_1' y_2' \vec{Q}_1' \vec{Q}_2' \left| \exp(iHt/\hbar) \right| y_1 y_2 \vec{Q}_1 \vec{Q}_2 \right\rangle \equiv K^*(y_1, y_2, \vec{Q}_1, \vec{Q}_2, t; y_1', y_2', \vec{Q}_1', \vec{Q}_2', 0) = \int \mathcal{D}\bar{y}_1 \mathcal{D}\bar{y}_2 \mathcal{D}\underline{\vec{Q}}_1 \mathcal{D}\underline{\vec{Q}}_2 \exp\{(-i/\hbar) S[y_1, y_2, \vec{Q}_1, \vec{Q}_2]\} \quad , \tag{5}$$

where the integration along all paths is carried out from $\bar{y}_1(0) = y_1'$ to $\bar{y}_1(t) = y_1$, from $\bar{y}_2(0) = y_2'$ to $\bar{y}_2(t) = y_2$, and from $\underline{\vec{Q}}_1(0) = Q_1'$ to $\underline{\vec{Q}}_1(t) = Q_1$, from $\underline{\vec{Q}}_2(0) = Q_2'$ to $\underline{\vec{Q}}_2(t) = Q_2$.

The action in Eqs.(4) and (5) is defined by

$$S(t) = \int_0^t \mathcal{L}(t') dt' , \tag{6}$$

where the Lagranjian is structured in according with Eq.(1)

$$\mathcal{L}(t) = \mathcal{L}_1(t) + \mathcal{L}_{B1}(t) + \mathcal{L}_{I1}(t) + \mathcal{L}_2(t) + \mathcal{L}_{B2}(t) + \mathcal{L}_{I2}(t) + \mathcal{L}_{12}(t) . \tag{7}$$

Taking into account Eqs.(1) and (7) the action in Eq.(6) can be written as follows

$$\begin{aligned}
S_i &= \int_0^t dt' \left[ M_i \dot{x}_i^2(t')/2 - M_i \omega_{0i}^2 x_i^2(t')/2 \right], \quad (i = 1, 2) \\
S_{12} &= \int_0^t dt' [\lambda x_1(t') x_2(t')], \\
S_{B1} &= \int_0^t dt' \sum_{j=1}^{N_1} m_j [\dot{q}_j^2(t')/2 - \omega_j^2 q_j^2(t')], \\
S_{B2} &= \int_0^t dt' \sum_{k=1}^{N_2} m_k [\dot{q}_k^2(t')/2 - \omega_k^2 q_k^2(t')], \\
S_{I1} &= \int_0^t dt' \sum_{j=1}^{N_1} [c_j q_j(t') x_1(t') - (c_j^2 / 2 m_j \omega_j^2) x_1^2(t')], \\
S_{I2} &= \int_0^t dt' \sum_{k=1}^{N_2} [c_k q_k(t') x_2(t') - (c_k^2 / 2 m_k \omega_k^2) x_2^2(t')]
\end{aligned} \quad . \tag{8}$$

In case $c_j = m_j \omega_j^2$ , $c_k = m_k \omega_k^2$ we obtain a system exactly corresponding to Hamiltonian in Eq. (1).

The reduced density matrix involving only variables of selected oscillators is obtained by tracing the whole density matrix in Eq.(3) over the variables of the two baths

$$\rho(x_1,x_2,y_1,y_2,t)=\int d\vec{R}_1 d\vec{R}_2 \left\langle x_1x_2\vec{R}_1\vec{R}_2 \left| W(t) \right| y_1y_2\vec{R}_1\vec{R}_2 \right\rangle$$
$$=\int dx_1'dx_2'dy_1'dy_2'd\vec{R}_1'd\vec{R}_2'd\vec{Q}_1'd\vec{Q}_2' \int d\vec{R}_1 d\vec{R}_2\, K(x_1,x_2,\vec{R}_1,\vec{R}_2,t;x_1',x_2',\vec{R}_1',\vec{R}_2',0)\,, \tag{9}$$
$$\times \left\langle x_1'x_2'\vec{R}_1'\vec{R}_2' \left| W(0) \right| y_1'y_2'\vec{Q}_1'\vec{Q}' \right\rangle K^*(y_1,y_2,\vec{R}_1,\vec{R}_2,t;y_1',y_2',\vec{Q}_1',\vec{Q}_2',0)$$

where

$$\left\langle x_1'x_2'\vec{R}_1'\vec{R}_2' \left| W(0) \right| y_1'y_2'\vec{Q}_1'\vec{Q}' \right\rangle \equiv W(x_1',x_2',y_1',y_2';\vec{R}_1',\vec{R}_2',\vec{Q}_1',\vec{Q}_2',0) \tag{10}$$

is the density matrix of the global system at the initial time $t=0$.

Our main goal in this paper is twofold: firstly, to represent analytically the reduced density matrix in Eq. (10) explicitly extracting the Feynman-Vernon influence functional and propagator for our case, and secondly to demonstrate an existence of the quasi-equilibrium state of the system under study.

## III. Results and discussion

### A. Feynman-Vernon influence functional and propagator for coupled oscillators

We choose the initial whole density matrix as follows

$$W(x_1',x_2',y_1',y_2';\vec{R}_1',\vec{R}_2',\vec{Q}_1',\vec{Q}_2',0)=\rho_A^{(1)}(x_1',y_1',0)\rho_A^{(2)}(x_2',y_2',0)\rho_B^{(1)}(\vec{R}_1',\vec{Q}_1',0)\rho_B^{(2)}(\vec{R}_2',\vec{Q}_2',0)\,, \tag{11}$$

where $\rho_A^{(1)}(x_1',y_1',0)$, $\rho_A^{(2)}(x_2',y_2',0)$ are initially prepared density matrices of the two selected oscillators, $\rho_B^{(1)}(\vec{R}_1',\vec{Q}_1',0)$, $\rho_B^{(2)}(\vec{R}_2',\vec{Q}_2',0)$ are initial density matrices of the separate reservoirs.

Then, the reduced density matrix looks like

$$\rho(x_1,x_2,y_1,y_2,t)=\int dx_1'dx_2'dy_1'dy_2'\; J(x_1,x_2,y_1,y_2,t;x_1',x_2',y_1',y_2',0)\,\rho_A^{(1)}(x_1',y_1',0)\rho_A^{(2)}(x_2',y_2',0)\,, \tag{12}$$

where the propagator

$$J(x_1,x_2,y_1,y_2,t;x_1',x_2',y_1',y_2',0)=\int \mathfrak{D}\overline{x}_1\mathfrak{D}\overline{x}_2\mathfrak{D}\overline{y}_1\mathfrak{D}\overline{y}_2\, \exp\left\{(i/\hbar)\left(S_{A1}[\overline{x}_1]-S_{A1}[\overline{y}_1] + S_{A2}[\overline{x}_2]-S_{A2}[\overline{y}_2]+S_{12}[\overline{x}_1,\overline{x}_2]-S_{12}[\overline{y}_1,\overline{y}_2]\right)\right\} \mathfrak{F}_{FV}[\overline{x}_1,\overline{x}_2,\overline{y}_1,\overline{y}_2]\,, \tag{13}$$

where the integration along all paths is carried out from $\overline{x}_1(0)=x_1'$ to $\overline{x}_1(t)=x_1$, from $\overline{x}_2(0)=x_2'$ to $\overline{x}_2(t)=x_2$, and from $\overline{y}_1(0)=y_1'$ to $\overline{y}_1(t)=y_1$, from $\overline{y}_2(0)=y_2'$ to $\overline{y}_2(t)=y_2$, correspondingly.

The Feynman-Vernon influence functional in Eq.(13) is equal to

$$\mathcal{F}_{FV}[\bar{x}_1,\bar{x}_2,\bar{y}_1,\bar{y}_2]=\int d\vec{R}_1' d\vec{R}_2' d\vec{Q}_1' d\vec{Q}_2' \int d\vec{R}_1 d\vec{R}_2\, \rho_B^{(1)}(\vec{R}_1',\vec{Q}_1',0)\rho_B^{(2)}(\vec{R}_2',\vec{Q}_2',0)$$
$$\int_{\vec{R}_1(0)=\vec{R}_1'}^{\vec{R}_1(t)=\vec{R}_1} \mathcal{D}\underline{\vec{R}}_1 \int_{\vec{R}_2(0)=\vec{R}_2'}^{\vec{R}_2(t)=\vec{R}_2} \mathcal{D}\underline{\vec{R}}_2 \int_{\vec{Q}_1(0)=\vec{Q}_1'}^{\vec{Q}_1(t)=\vec{R}_1} \mathcal{D}\underline{\vec{Q}}_1 \int_{\vec{Q}_2(0)=\vec{Q}_2'}^{\vec{Q}_2(t)=\vec{R}_2} \mathcal{D}\underline{\vec{Q}}_2 \exp\Big\{(i/\hbar)\Big(S_{I1}[\bar{x}_1,\underline{\vec{R}}_1]-S_{I1}[\bar{y}_1,\underline{\vec{Q}}_1] \quad . \tag{14}$$
$$+S_{I2}[\bar{x}_2,\underline{\vec{R}}_2]-S_{I2}[\bar{y}_2,\underline{\vec{Q}}_2]+S_{B1}[\underline{\vec{R}}_1]-S_{B1}[\underline{\vec{Q}}_1]+S_{B2}[\underline{\vec{R}}_2]-S_{B2}[\underline{\vec{Q}}_2]\Big)\Big\}$$

It is clear from Eq. (14) that this influence functional can be represented in to the factorizable form

$$\mathcal{F}_{FV}[\bar{x}_1,\bar{x}_2,\bar{y}_1,\bar{y}_2]=\mathcal{F}_{FV}^{(1)}[\bar{x}_1,\bar{y}_1]\mathcal{F}_{FV}^{(2)}[\bar{x}_2,\bar{y}_2], \tag{15}$$

where

$$\mathcal{F}_{FV}^{(1)}[\bar{x}_1,\bar{y}_1]=\int d\vec{R}_1' d\vec{Q}_1' \int d\vec{R}_1\, \rho_B^{(1)}(\vec{R}_1',\vec{Q}_1',0)$$
$$\int_{\vec{R}_1(0)=\vec{R}_1'}^{\vec{R}_1(t)=\vec{R}_1} \mathcal{D}\underline{\vec{R}}_1 \int_{\vec{Q}_1(0)=\vec{Q}_1'}^{\vec{Q}_1(t)=\vec{R}_1} \mathcal{D}\underline{\vec{Q}}_1 \exp\left\{\frac{i}{\hbar}\Big(S_{I1}[\bar{x}_1,\underline{\vec{R}}_1]-S_{I1}[\bar{y}_1,\underline{\vec{Q}}_1]+S_{B1}[\underline{\vec{R}}_1]-S_{B1}[\underline{\vec{Q}}_1]\Big)\right\}, \tag{16}$$

and

$$\mathcal{F}_{FV}^{(2)}[\bar{x}_1,\bar{y}_1]=\int d\vec{R}_2' d\vec{Q}_2' \int d\vec{R}_2\, \rho_B^{(2)}(\vec{R}_2',\vec{Q}_2',0)$$
$$\int_{\vec{R}_2(0)=\vec{R}_2'}^{\vec{R}_2(t)=\vec{R}_2} \mathcal{D}\underline{\vec{R}}_2 \int_{\vec{Q}_2(0)=\vec{Q}_2'}^{\vec{Q}_2(t)=\vec{R}_2} \mathcal{D}\underline{\vec{Q}}_2 \exp\left\{\frac{i}{\hbar}\Big(S_{I2}[\bar{x}_2,\underline{\vec{R}}_2]-S_{I2}[\bar{y}_2,\underline{\vec{Q}}_2]+S_{B2}[\underline{\vec{R}}_2]-S_{B2}[\underline{\vec{Q}}_2]\Big)\right\}, \tag{17}$$

Such path integrals are evaluated previously, see for instance [10, 45]. In our case we have

$$\mathcal{F}_{FV}[\bar{x}_1,\bar{x}_2,\bar{y}_1,\bar{y}_2]=\exp\Big\{-\Big(\Phi_{EV}^{(1)}[\bar{x}_1,\bar{y}_1]+\Phi_{EV}^{(2)}[\bar{x}_2,\bar{y}_2]\Big)\Big\}, \tag{18}$$

$$\begin{aligned}\Phi_{EV}^{(1)}[\bar{x}_1,\bar{y}_1]+\Phi_{EV}^{(2)}[\bar{x}_2,\bar{y}_2]&=\hbar^{-1}\int_0^t dt'\int_0^{t'} dt''\{\bar{x}_1(t')-\bar{y}_1(t')\}\{\alpha_1(t'-t'')\bar{x}_1(t'')-\alpha_1^*(t'-t'')\bar{y}_1(t'')\}\\
&+i(\mu_1/2\hbar)\int_0^t dt'[\bar{x}_1^2(t')-\bar{y}_1^2(t')]+\\
&\hbar^{-1}\int_0^t dt'\int_0^{t'} dt''\{\bar{x}_2(t')-\bar{y}_2(t')\}\{\alpha_2(t'-t'')\bar{x}_2(t'')-\alpha_2^*(t'-t'')\bar{y}_2(t'')\}\\
&+i(\mu_2/2\hbar)\int_0^t dt'[\bar{x}_2^2(t')-\bar{y}_2^2(t')]\end{aligned}, \tag{19}$$

where

$$\mu_1=\sum_{j=1}^N\frac{c_j^2}{m_j\omega_j^2}=\frac{2}{\pi}\int_0^\infty d\omega\,\frac{J_1(\omega)}{\omega},\quad \mu_2=\sum_{k=1}^N\frac{c_k^2}{m_k\omega_k^2}=\frac{2}{\pi}\int_0^\infty d\omega\,\frac{J_2(\omega)}{\omega}, \tag{20}$$

where $J_{1,2}(\omega)$ is the spectral densities of noise of two baths,

$$\alpha_1(z)=\sum_{j=1}^N\frac{c_j^2}{m_j\omega_j}\frac{\cosh[\omega_j(\hbar\beta_1/2-iz)]}{\sinh[\omega_j\hbar\beta_1/2]}=\frac{1}{\pi}\int_0^\infty d\omega\,J_1(\omega)\frac{\cosh[\omega(\hbar\beta_1/2-iz)]}{\sinh[\omega\hbar\beta_1/2]}, \tag{21}$$

and similar expressions for $\alpha_2$.

The functions $\alpha_{1,2}$ can be represented in the following form

$$\alpha_{1,2}(t) \equiv \alpha'_{1,2}(t) + i\alpha''_{1,2}(t) = \frac{1}{\pi}\int_0^\infty d\omega\, J_{1,2}(\omega)[\coth(\omega\hbar\beta_{1,2}/2)\cos(\omega t) - i\sin(\omega t)]. \qquad (22)$$

Taking into account Eq. (22) the Eq. (13) for propagator can be rewritten as follows

$$\begin{aligned}
&J(x_1,x_2,y_1,y_2,t;x'_1,x'_2,y'_1,y'_2,0) = \\
&\int \mathcal{D}\overline{x}_1\mathcal{D}\overline{y}_1 \exp\frac{i}{\hbar}\left\{S_{A1}[\overline{x}_1] - S_{A1}[\overline{y}_1] - \int_0^t dt'\int_0^{t'} dt''\{\overline{x}_1(t') - \overline{y}_1(t')\}\alpha''_1(t'-t'')\{\overline{x}_1(t'') + \overline{y}_1(t'')\}\right\} \\
&\times \exp-\frac{1}{\hbar}\int_0^t dt'\int_0^{t'} dt''\{\overline{x}_1(t') - \overline{y}_1(t')\}\alpha'_1(t'-t'')\{\overline{x}_1(t'') - \overline{y}_1(t'')\} \exp-i\frac{\mu_1}{2\hbar}\int_0^t dt'[\overline{x}_1^2(t') - \overline{y}_1^2(t')] \\
&\times\int \mathcal{D}\overline{x}_2\mathcal{D}\overline{y}_2 \exp\frac{i}{\hbar}\{S_{A2}[\overline{x}_2] - S_{A2}[\overline{y}_2] + S_{12}[\overline{x}_1,\overline{x}_2] - S_{12}[\overline{y}_1,\overline{y}_2] - \\
&-\int_0^t dt'\int_0^{t'} dt''\{\overline{x}_2(t') - \overline{y}_2(t')\}\alpha''_2(t'-t'')\{\overline{x}_2(t'') + \overline{y}_2(t'')\} \\
&\times \exp-\frac{1}{\hbar}\int_0^t dt'\int_0^{t'} dt''\{\overline{x}_2(t') - \overline{y}_2(t')\}\alpha'_2(t'-t'')\{\overline{x}_2(t'') - \overline{y}_2(t'')\} \exp-i\frac{\mu_2}{2\hbar}\int_0^t dt'[\overline{x}_2^2(t') - \overline{y}_2^2(t')]
\end{aligned} . \qquad (23)$$

Here we use the approximation accepted in [45], considering the Ohmic case $J_{1,2}(\omega) = \eta_{1,2}\omega$, which gives

$L''_{1,2}(t) \approx \eta_{1,2}\partial\delta(t)/\partial t$, where $\delta(t)$ is the delta-function, $\mu_{1,2} = 2\eta_{1,2}\nu_{1,2}^{\max}/\pi$, where $\nu_{1,2}^{\max}$ is the maximal frequency of excitations within baths. In this case we obtain from Eq. (23)

$$\begin{aligned}
&J(x_1,x_2,y_1,y_2,t;x'_1,x'_2,y'_1,y'_2,0) = \int \mathcal{D}\overline{x}_1\mathcal{D}\overline{y}_1\mathcal{D}\overline{x}_2\mathcal{D}\overline{y}_2 \exp\frac{i}{\hbar}\{S_{A1}[\overline{x}_1] - S_{A1}[\overline{y}_1] \\
&+S_{A2}[\overline{x}_2] - S_{A2}[\overline{y}_2] + S_{12}[\overline{x}_1,\overline{x}_2] - S_{12}[\overline{y}_1,\overline{y}_2] \\
&-M_1\gamma_1\int_0^t dt'[\overline{x}_1(t')\dot{\overline{x}}_1(t') - \overline{y}_1(t')\dot{\overline{y}}_1(t') + \overline{x}_1(t')\dot{\overline{y}}_1(t') - \overline{y}_1(t')\dot{\overline{x}}_1(t')] \\
&-M_2\gamma_2\int_0^t dt'[\overline{x}_2(t')\dot{\overline{x}}_2(t') - \overline{y}_2(t')\dot{\overline{y}}_2(t') + \overline{x}_2(t')\dot{\overline{y}}_2(t') - \overline{y}_2(t')\dot{\overline{x}}_2(t')]\Big\} \exp[-(\phi_1 + \phi_2)]
\end{aligned} , \qquad (24)$$

where we desigtated $\gamma_{1,2} = \eta_{1,2}/2M_{1,2}$, and $\phi_i \equiv \phi_i[\overline{x}_i,\overline{y}_i]$ are equal

$$\phi_i = \frac{2M_i\gamma_i}{\hbar\pi}\int_0^{\nu_i^{\max}} d\omega\,\omega\, Coth\left(\frac{\hbar\omega}{2k_BT_i}\right)\int_0^t dt'\int_0^{t'} dt''\{\overline{x}_i(t') - \overline{y}_i(t')\}\cos[\omega(t'-t'')]\{\overline{x}_i(t'') - \overline{y}_i(t'')\}, \qquad (25)$$

where $i = 1, 2$.

The Eq. (24) for propagator can be again rewritten as follows

$$J(x_1,x_2,y_1,y_2,t;x'_1,x'_2,y'_1,y'_2,0) = \int \mathcal{D}\overline{x}_1\mathcal{D}\overline{y}_1\mathcal{D}\overline{x}_2\mathcal{D}\overline{y}_2 \exp\frac{i}{\hbar}S[\overline{x}_1,\overline{x}_2,\overline{y}_1,\overline{y}_2]\exp[-(\phi_1+\phi_2)], \qquad (26)$$

where

$$S[\overline{x}_1,\overline{x}_2,\overline{y}_1,\overline{y}_2] = \int_0^t dt'\,\mathcal{L}(\overline{x}_1,\overline{x}_2,\dot{\overline{x}}_1,\dot{\overline{x}}_2,\overline{y}_1,\overline{y}_2,\dot{\overline{y}}_1,\dot{\overline{y}}_2). \qquad (27)$$

Now, we consider the Lagrangian from Eq. (27) omitting the straight lines over the letters for brevity

$$\begin{aligned}\mathcal{L} &= M_1\dot{x}_1^2/2 - M_1\dot{y}_1^2/2 + M_2\dot{x}_2^2/2 - M_2\dot{y}_2^2/2 \\ &-M_1\omega_{01}^2 x_1^2/2 + M_1\omega_{01}^2 y_1^2/2 - M_2\omega_{02}^2 x_2^2/2 + M_2\omega_{02}^2 y_2^2/2 \\ &-M_1\gamma_1[x_1\dot{x}_1 - y_1\dot{y}_1 + x_1\dot{y}_1 - y_1\dot{x}_1] - M_2\gamma_2[x_2\dot{x}_2 - y_2\dot{y}_2 + x_2\dot{y}_2 - y_2\dot{x}_2] \\ &+\lambda x_1 x_2 - \lambda y_1 y_2\end{aligned} \quad . \qquad (28)$$

The Lagrange equations of motion

$$\frac{d}{dt}\frac{\partial \mathcal{L}}{\partial \dot{x}_i} - \frac{\partial \mathcal{L}}{\partial x_i} = 0, \quad \frac{d}{dt}\frac{\partial \mathcal{L}}{\partial \dot{y}_i} - \frac{\partial \mathcal{L}}{\partial y_i} = 0, \quad (i = 1, 2), \qquad (29)$$

as applied to the Lagrangian in Eq. (28) give the following system of coupled equations

$$\begin{cases} \ddot{x}_1 + 2\gamma_1\dot{y}_1 + \omega_{01}^2 x_1 = (\lambda / M_1) x_2 \\ \ddot{x}_2 + 2\gamma_2\dot{y}_2 + \omega_{02}^2 x_2 = (\lambda / M_2) x_1 \\ \ddot{y}_1 + 2\gamma_1\dot{x}_1 + \omega_{01}^2 y_1 = (\lambda / M_1) y_2 \\ \ddot{y}_2 + 2\gamma_2\dot{x}_2 + \omega_{02}^2 y_2 = (\lambda / M_2) y_1 \end{cases} . \qquad (30)$$

Solution of the system is done in Appendix A. The classical solution is obtained in new variables $X_{1,2} = x_{1,2} + y_{1,2}$, $\xi_{1,2} = x_{1,2} - y_{1,2}$. In these new variables the Lagrangian in Eq. (28) reads

$$\begin{aligned}\mathcal{L} &= M_1\dot{X}_1\dot{\xi}_1/2 - M_1\omega_{01}^2 X_1\xi_1/2 - M_1\gamma_1\dot{X}_1\xi_1 \\ &+M_2\dot{X}_2\dot{\xi}_2/2 - M_2\omega_{02}^2 X_2\xi_2/2 - M_2\gamma_2\dot{X}_2\xi_2 \\ &+(\lambda/2)(X_1\xi_2 + X_2\xi_1)\end{aligned} \quad . \qquad (31)$$

Now, as well as in [45, 46] we represent the paths as the sums $X_{1,2} = \tilde{X}_{1,2} + X'_{1,2}$, $\xi_{1,2} = \tilde{\xi}_{1,2} + \xi'_{1,2}$, explicitly selecting classical paths in Eqs.(A8)-(A11) and fluctuating parts $X'_{1,2}$, $\xi'_{1,2}$ with boundary conditions $X'_{1,2}(0) = X'_{1,2}(t) = 0$, $\xi'_{1,2}(0) = \xi'_{1,2}(t) = 0$. The Lagrangian in Eq.(31) becomes

$$\begin{aligned}\mathcal{L} \equiv \tilde{\mathcal{L}} + \mathcal{L}' &= M_1\dot{\tilde{X}}_1\dot{\tilde{\xi}}_1/2 - M_1\omega_{01}^2\tilde{X}_1\tilde{\xi}_1/2 - M_1\gamma_1\dot{\tilde{X}}_1\tilde{\xi}_1 \\ &+M_2\dot{\tilde{X}}_2\dot{\tilde{\xi}}_2/2 - M_2\omega_{02}^2\tilde{X}_2\tilde{\xi}_2/2 - M_2\gamma_2\dot{\tilde{X}}_2\tilde{\xi}_2 + (\lambda/2)(\tilde{X}_1\tilde{\xi}_2 + \tilde{X}_2\tilde{\xi}_1) \\ &+M_1\dot{X}'_1\dot{\xi}'_1/2 - M_1\omega_{01}^2 X'_1\xi'_1/2 - M_1\gamma_1\dot{X}'_1\xi'_1 \\ &+M_2\dot{X}'_2\dot{\xi}'_2/2 - M_2\omega_{02}^2 X'_2\xi'_2/2 - M_2\gamma_2\dot{X}'_2\xi'_2 + (\lambda/2)(X'_1\xi'_2 + X'_2\xi'_1)\end{aligned} \quad . \qquad (32)$$

Here we would like to give the essential point in our consideration. This paper addresses to solve the important problem of a weak coupling $\lambda \to 0$ between two selected oscillators. In this case the eigenfrequencies in Eq.(A4) of the system can be represented as follows

$$\Omega_{1,2}^2 = \omega_{01,02}^2 \mp \lambda^2 / M_1 M_2(\omega_{02}^2 - \omega_{01}^2) . \qquad (33)$$

It is well known that together with $r_1$ and $r_2$ parameters in Eq. (A6) the coupling between two interacting oscillators can be characterized by the coupling coefficient

$$\rho = \frac{\lambda}{\sqrt{M_1 M_2}\,\omega_{01}\omega_{02}} \ . \tag{34}$$

In case of a weak coupling $\rho \ll 1$ when the new eigenfrequencies can be expressed by Eq.(33) we have

$$r_1 = \frac{\lambda}{M_2(\omega_{02}^2 - \omega_{01}^2)} \ , \quad r_1 = -\frac{\lambda}{M_1(\omega_{02}^2 - \omega_{01}^2)} \ . \tag{35}$$

In this approach it is follows from Eq.(A5) that

$$\delta_1 = \gamma_1 + \frac{\lambda^2 \gamma_2}{M_1 M_2 (\omega_{02}^2 - \omega_{01}^2)^2} \ , \quad \delta_2 = \gamma_2 - \frac{\lambda^2 \gamma_1}{M_1 M_2 (\omega_{02}^2 - \omega_{01}^2)^2} \, . \tag{36}$$

In our further calculations we neglect terms containing $r_1^n$ and $r_2^n$ with $n \geq 2$ . In other words, we seek for a solution in the linear regime with respect to $\lambda$ . Thus, taking into account Eqs.(33),(34) we can put for simplicity $\Omega_{1,2} \approx \omega_{01,02}$ and $\delta_{1,2} \approx \gamma_{1,2}$ in this approach. In all our further calculations we have used parameters of both oscillators which satisfy the inequality $(\omega_{02}^2 - \omega_{01}^2)/2\omega_{01}\omega_{02} \gg \rho$ , manifesting the weak coupling in the system. It should be emphasized that the special case of identical oscillators takes a separate consideration because $r_1 = 1$ , $r_2 = -1$ and no small parameter is involved.

With the above described assumptions and taking into account Eq.(32) the action in Eq.(27) calculated in new variables Eqs.(A8)-(A11) is expressed as follows

$$\tilde{S}_{cl} = \int_0^t \tilde{\mathcal{L}}(t')dt' = \tilde{S}_{cl}^{(1)} + \tilde{S}_{cl}^{(2)} + \tilde{S}_{cl}^{(12)}, \quad S' = \int_0^t \mathcal{L}'(t')dt' = S_1' + S_2' + S_{12}' \, , \tag{37}$$

where

$$\begin{aligned} \tilde{S}_{cl}^{(1)} + \tilde{S}_{cl}^{(2)} = {} & D_1(X_{f1}\xi_{f1}) + [D_5 + D_5'](X_{f2}\xi_{f1}) + [D_6 + D_6'](X_{f1}\xi_{f2}) + D_1'(X_{f2}\xi_{f2}) + \\ & + D_2(X_{i1}\xi_{f1}) + D_3(X_{f1}\xi_{i1}) + D_4(X_{i1}\xi_{i1}) + \\ & + D_2'(X_{i2}\xi_{f2}) + D_3'(X_{f2}\xi_{i2}) + D_4'(X_{i2}\xi_{i2}) + \\ & + [D_7 + D_7'](X_{i1}\xi_{f2}) + [D_8 + D_8'](X_{i2}\xi_{f1}) + [D_9 + D_9'](X_{f1}\xi_{i2}) + [D_{10} + D_{10}'](X_{f2}\xi_{i1}) + \\ & + [D_{11} + D_{11}'](X_{i1}\xi_{i2}) + [D_{12} + D_{12}'](X_{i2}\xi_{i1}) \end{aligned} \ , \tag{38}$$

where temporal functions $D_k \equiv D_k(t)$ and $D_k' \equiv D_k'(t)$ $k = 1,...,12$ are presented in Appendix B,

$$\begin{aligned} \tilde{S}_{cl}^{(12)} = {} & \Pi_1(X_{f1}\xi_{f1}) + \Pi_2(X_{f1}\xi_{f2}) + \Pi_3(X_{f2}\xi_{f1}) + \Pi_4(X_{f2}\xi_{f2}) + \\ & + \Pi_5(X_{f1}\xi_{i1}) + \Pi_6(X_{f1}\xi_{i2}) + \Pi_7(X_{f2}\xi_{i1}) + \Pi_8(X_{f2}\xi_{i2}) + \\ & + \Pi_9(X_{i1}\xi_{f1}) + \Pi_{10}(X_{i1}\xi_{f2}) + \Pi_{11}(X_{i2}\xi_{f1}) + \Pi_{12}(X_{i2}\xi_{f2}) + \\ & + \Pi_{13}(X_{i1}\xi_{i1}) + \Pi_{14}(X_{i1}\xi_{i2}) + \Pi_{15}(X_{i2}\xi_{i1}) + \Pi_{16}(X_{i2}\xi_{i2}) \end{aligned} \ , \tag{39}$$

where temporal functions $\Pi_k \equiv \Pi_k(t)$ , $k = 1,...,16$ are written in Appendix B.

After calculations of the classical actions and functions $\phi_{1,2}$ in Eq.(24) one gets the final expression for the propagator in case of two coupled oscillators

$$
\begin{aligned}
&J(X_{f1},X_{f2},\xi_{f1},\xi_{f2},t;X_{i1},X_{i2},\xi_{i1},\xi_{i2},0)=\exp\frac{i}{\hbar}\{\tilde{S}_{cl}^{(1)}+\tilde{S}_{cl}^{(2)}+\tilde{S}_{cl}^{(12)}\}\\
&\times\exp-\frac{1}{\hbar}\{A_1(t)\xi_{f1}^2+B_1(t)\xi_{f1}\xi_{i1}+C_1(t)\xi_{i1}^2\}\exp-\frac{1}{\hbar}\{A_2(t)\xi_{f2}^2+B_2(t)\xi_{f2}\xi_{i2}+C_2(t)\xi_{i2}^2\}\\
&\times\exp-\frac{1}{\hbar}\{E_1(t)\xi_{i1}\xi_{i2}+E_2(t)\xi_{f2}\xi_{i1}+E_3(t)\xi_{f1}\xi_{i2}+E_4(t)\xi_{f1}\xi_{f2}\}\\
&\times G(X_{f1},X_{f2},\xi_{f1},\xi_{f2},t;X_{i1},X_{i2},\xi_{i1},\xi_{i2},0)
\end{aligned}\quad,\qquad (40)
$$

where

$$A_i(t)=\frac{M_i\gamma_i}{\pi}\int_0^{\nu_i^{\max}}d\omega\,\omega\,Coth\left(\frac{\hbar\omega}{2k_BT_i}\right)A_\omega^{(i)}(t),\ \ (i=1,2)\,,\qquad (41)$$

$$A_\omega^{(i)}(t)=\frac{\exp(-2\delta_it)}{Sin^2(\Omega_it)}\int_0^t\int_0^t dt'dt''\,Sin(\Omega_it')Cos[\omega(t'-t'')]Sin(\Omega_it'')\exp[\delta_i(t'+t'')],\ \ (i=1,2)\,,\qquad (42)$$

$$B_i(t)=\frac{M_i\gamma_i}{\pi}\int_0^{\nu_i^{\max}}d\omega\,\omega\,Coth\left(\frac{\hbar\omega}{2k_BT_i}\right)B_\omega^{(i)}(t),\ \ (i=1,2)\,,\qquad (43)$$

$$B_\omega^{(i)}(t)=\frac{2\exp(-\delta_it)}{Sin^2(\Omega_it)}\int_0^t\int_0^t dt'dt''\,Sin(\Omega_it')Cos[\omega(t'-t'')]Sin[\Omega_i(t-t'')]\exp[\delta_i(t'+t'')],\ \ (i=1,2)\,,\qquad (44)$$

$$C_i(t)=\frac{M_i\gamma_i}{\pi}\int_0^{\nu_i^{\max}}d\omega\,\omega\,Coth\left(\frac{\hbar\omega}{2k_BT_i}\right)C_\omega^{(i)}(t),\ \ (i=1,2)\,,\qquad (45)$$

$$C_\omega^{(i)}(t)=\frac{1}{Sin^2(\Omega_it)}\int_0^t\int_0^t dt'dt''\,Sin[\Omega_i(t-t')]Cos[\omega(t'-t'')]Sin[\Omega_i(t-t'')]\exp[\delta_i(t'+t'')],\ \ (i=1,2)\,,\qquad (46)$$

time dependent functions $E_{1,2,3,4}(t)$ are written in Appendix D.

The fluctuational integral in Eq.(40) is

$$
\begin{aligned}
&G(X_{f1},X_{f2},\xi_{f1},\xi_{f2},t;X_{i1},X_{i2},\xi_{i1},\xi_{i2},0)=\frac{1}{4}\int\mathfrak{D}X_1'\mathfrak{D}X_2'\mathfrak{D}\xi_1'\mathfrak{D}\xi_2'\exp\frac{i}{\hbar}\{S_1'+S_2'+S_{12}'\}\\
&\times\exp-\frac{2}{\hbar}\{\phi_{T1}[\xi_1',\xi_1']+\phi_{T2}[\xi_2',\xi']\}\exp-\frac{1}{\hbar}\{\phi_{T1}[\tilde{\xi}_1,\xi_1']+\phi_{T2}[\tilde{\xi}_2,\xi_2']\}
\end{aligned}\quad,\qquad (47)
$$

where the integration is carried out along all closed paths because $X_{1,2}'(0)=X_{1,2}'(t)=0$ and $\xi_{1,2}'(0)=\xi_{1,2}'(t)=0$,

$$\phi_{Ti}[a,b]=\frac{M_i\gamma_i}{\hbar\pi}\int_0^{\nu_i^{\max}}d\omega\,\omega\,Coth\left(\frac{\hbar\omega}{2k_BT_i}\right)\int_0^t dt'\int_0^t dt''\,a(t')\cos[\omega(t'-t'')]b(t''),\ \ (i=1,2)\,,\qquad (48)$$

The fluctuational integral in Eq.(47) is calculated in Appendix D. This is exactly the square of the product of the wave functions amplitudes [45,46] of two uncoupled oscillators for the undumped case

$$G(X_{f1}, X_{f2}, \xi_{f1}, \xi_{f2}, t; X_{i1}, X_{i2}, \xi_{i1}, \xi_{i2}, 0) = \tilde{C}_1 \tilde{C}_2 F_1^2(t) F_2^2(t) . \qquad (49)$$

### B. Reduced density matrix of coupled oscillators

Thus, taking into account Eq.(49) the final form of the propagating function in Eq.(40) reads

$$\begin{aligned} &J(X_{f1}, X_{f2}, \xi_{f1}, \xi_{f2}, t; X_{i1}, X_{i2}, \xi_{i1}, \xi_{i2}, 0) = \tilde{C}_1 \tilde{C}_2 F_1^2(t) F_2^2(t) \exp \frac{i}{\hbar} \tilde{S}_{cl}^{(12)} \\ &\times \exp \frac{i}{\hbar} \tilde{S}_{cl}^{(1)} \exp - \frac{1}{\hbar} \{ A_1(t) \xi_{f1}^2 + B_1(t) \xi_{f1} \xi_{i1} + C_1(t) \xi_{i1}^2 \} \\ &\times \exp \frac{i}{\hbar} \tilde{S}_{cl}^{(2)} \exp - \frac{1}{\hbar} \{ A_2(t) \xi_{f2}^2 + B_2(t) \xi_{f2} \xi_{i2} + C_2(t) \xi_{i2}^2 \} \\ &\times \exp - \frac{1}{\hbar} \{ E_1(t) \xi_{i1} \xi_{i2} + E_2(t) \xi_{f2} \xi_{i1} + E_3(t) \xi_{f1} \xi_{i2} + E_4(t) \xi_{f1} \xi_{f2} \} \end{aligned} , \qquad (50)$$

which means that the dampings modify only exponents of as well as in [45] and, additionally, the coupling between two oscillators is described just the separate exponents in Eq.(50) .

In order to calculate the reduced density matrix $\rho(x_1, x_2, y_1, y_2, t)$ in Eq.(12) we ought to assign the initial density matrixes $\rho_A^{(1)}(x_1', y_1', 0)$ and $\rho_A^{(2)}(x_2', y_2', 0)$ for two selected oscillators. In new variables we have instead of Eq.(12) the following expression for propagator

$$\begin{aligned} \rho(X_{f1}, X_{f2}, \xi_{f1}, \xi_{f2}, t) = \int dX_{i1} dX_{i2} d\xi_{i1} d\xi_{i2} \; & J(X_{f1}, X_{f2}, \xi_{f1}, \xi_{f2}, t; X_{i1}, X_{i2}, \xi_{i1}, \xi_{i2}, 0) \\ & \times \rho_A^{(1)}(X_{i1}, \xi_{i1}, 0) \rho_A^{(2)}(X_{i2}, \xi_{i2}, 0) \end{aligned} , \qquad (51)$$

where we choose for illustrative purpose

$$\begin{aligned} \rho_A^{(1)}(X_{i1}, \xi_{i1}, 0) &= (2\pi\sigma_{01}^2)^{-1/2} \exp\left[ -(X_{i1}^2 + \xi_{i1}^2)/8\sigma_{01}^2 \right], \\ \rho_A^{(2)}(X_{i2}, \xi_{i2}, 0) &= (2\pi\sigma_{02}^2)^{-1/2} \exp\left[ -(X_{i2}^2 + \xi_{i2}^2)/8\sigma_{02}^2 \right], \end{aligned} \qquad (52)$$

where $\sigma_{01}^2 = \hbar / 2M_1\omega_{01}$ and $\sigma_{02}^2 = \hbar / 2M_2\omega_{02}$ are the initial spatial variances of two oscillators.

An integration in Eq.(51) is straghtforward using Eqs.(50),(51) but very tedious and leads to a cumbersome result. Here, following to [45] we consider a simpler case when $X_f = 2x_f$ and $\xi_f = 0$. In this case we can represent the density matrix in Eq.(51) after an integration as follows

$$\rho(x_{f1}, x_{f2}, t) = \rho_{01}(t) \rho_{02}(t) \exp\left[ -\frac{x_{f1}^2}{2\sigma_1^2(t)} - \frac{x_{f1} x_{f2}}{\beta_{12}(t)} - \frac{x_{f2}^2}{2\sigma_2^2(t)} \right], \qquad (53)$$

where

$$\sigma_1^2(t,T_1)=\frac{1}{2}\left\{\frac{D_3^2(t)}{\hbar[C_1(t)+\hbar a_1]}\left[1-\frac{D_4^2(t)}{D_4^2(t)+4\hbar a_1[C_1(t)+\hbar a_1]}\right]\right\}^{-1}, \quad (54)$$

$$\sigma_2^2(t,T_2)=\frac{1}{2}\left\{\frac{D_3'^2(t)}{\hbar[C_2(t)+\hbar a_2]}\left[1-\frac{D_4'^2(t)}{D_4'^2(t)+4\hbar a_2[C_2(t)+\hbar a_2]}\right]\right\}^{-1}, \quad (55)$$

$$\beta_{12}(t,\lambda,T_1,T_2)=\left\{\frac{2D_3'(t)f_1(t)}{\hbar[C_2(t)+\hbar a_2]}\left[1-\frac{D_4'^2(t)}{D_4'^2(t)+4a_2\hbar[C_2(t)+\hbar a_2]}\right]+\right.$$
$$\left.+\frac{2D_3(t)f_2(t)}{\hbar[C_1(t)+\hbar a_1]}\left[1-\frac{D_4^2(t)}{D_4^2(t)+4a_1\hbar[C_1(t)+\hbar a_1]}\right]\right\}^{-1}, \quad (56)$$

$$\rho_{01}(t)\rho_{02}(t)=\frac{\tilde{C}_1F_1^2(t)}{\sqrt{C_1(t)+\hbar a_1+D_4^2(t)/4\hbar a_1}}\frac{\tilde{C}_2F_2^2(t)}{\sqrt{C_2(t)+\hbar a_2+D_4'^2(t)/4\hbar a_2}}, \quad (57)$$

where $a_1=1/8\sigma_{01}^2$, $a_2=1/8\sigma_{02}^2$ and the time dependent functions are presented in Eqs.(45),(46) and in Appendices B and D.

First of all, it should be emphasized that in the described approach of weak coupling or in the linear regime with respect to $\lambda$ the variances $\sigma_1^2(t,T_1)$ and $\sigma_2^2(t,T_2)$ are not depended on this coupling constant at all. Each spatial variance depends only on the temperature of the own thermostat. Contrary to these variances the covariance $\beta_{12}(t,\lambda,T_1,T_2)$ depends both on temperatures of two thermostats and on the coupling parameter $\lambda$.

It is easy to demonstrate analytical asymptotes of $\sigma_1^2(t,T_1)$ and $\sigma_2^2(t,T_2)$ at $t\to 0$ and at $t\to\infty$.

For example, we have from Eqs.(45),(46) and Eqs.(B3),(B4) the following limiting temporal behavior of the functions

$$\begin{aligned}&D_3(t\to 0)\to -M_1/2t, \quad D_3(t\to\infty)\to -M_1\omega_1\exp(\gamma_1 t)/\sin(\omega_1 t),\\ &D_4(t\to 0)\to M_1/2t, \quad D_4(t\to\infty)\to M_1\omega_1/2,\end{aligned} \quad (58)$$

$$C_1(t\to 0)\to 0, \quad C_1(t\to\infty)\to\frac{M_1^2\omega_1^2\exp(2\gamma_1 t)}{2\hbar\sin^2(\omega_1 t)}\sigma_1^2(FDT), \quad (59)$$

$$\sigma_1^2(FDT)=\frac{\hbar}{\pi}\int_0^\infty d\omega\, Coth\left(\frac{\hbar\omega}{2k_BT_1}\right)\frac{2\gamma_1\omega}{M_1[(\omega^2-\omega_{01}^2)^2+4\gamma_1^2\omega^2]}, \quad (60)$$

and similar relations for $D_3'$, $D_4'$, $C_2$ and for $\sigma_2^2(FDT)$ with characteristics of the second oscillator and temperature $T_2$.

In this two limiting cases we obtain at $t\to 0$

$$\rho(x_{f1}, t \to 0) \to [2\pi\sigma_1^2(0)]^{-1/2} \exp\left[-x_{f1}^2 / 2\sigma_1^2(0)\right],$$
$$\rho(x_{f2}, t \to 0) \to [2\pi\sigma_2^2(0)]^{-1/2} \exp\left[-x_{f2}^2 / 2\sigma_2^2(0)\right], \tag{61}$$

as it must be because we assigned Eq.(52) at $t = 0$, where $\sigma_1^2(0) \equiv \sigma_{01}^2 = \hbar / 2M_1\omega_{01}$ and $\sigma_2^2(0) \equiv \sigma_{02}^2 = \hbar / 2M_2\omega_{02}$ are the initial spatial variances of two oscillators.

And at $t \to \infty$

$$\rho(x_{f1}, t \to \infty) \to [2\pi\sigma_1^2(FDT)]^{-1/2} \exp\left[-x_{f1}^2 / 2\sigma_1^2(FDT)\right],$$
$$\rho(x_{f2}, t \to \infty) \to [2\pi\sigma_2^2(FDT)]^{-1/2} \exp\left[-x_{f2}^2 / 2\sigma_2^2(FDT)\right], \tag{62}$$

where $\sigma_1^2(FDT)$ from Eq.(60) and $\sigma_2^2(FDT)$ is given by the similar formula.

It is interesting to compare the above obtained Eqs.(61) and (62) with the textbook formula [53]

$$\rho(x) = \left[\frac{M\omega}{\pi\hbar} Tanh\left(\frac{\hbar\omega}{2k_B T}\right)\right]^{1/2} \exp\left[-\frac{M\omega x^2}{\hbar} Tanh\left(\frac{\hbar\omega}{2k_B T}\right)\right]. \tag{63}$$

At $T = 0K$ we obtain from Eq.(63) of course Eqs.(52) and (61) because this is exactly our initial conditions.

At any finite temperature in Eq.(63) we can to write

$$\frac{\pi\hbar}{M\omega} Coth\left(\frac{\hbar\omega}{2k_B T}\right) = \frac{2\pi\hbar}{M}\int_0^\infty d\nu\, Coth\left(\frac{\hbar\nu}{2k_B T}\right)\delta(\nu^2 - \omega^2)$$
$$= \frac{2\hbar}{M}\lim_{\varepsilon \to 0}\int_0^\infty d\nu\, Coth\left(\frac{\hbar\nu}{2k_B T}\right)\frac{\varepsilon}{(\nu^2 - \omega^2)^2 + \varepsilon^2}, \tag{64}$$

where $\delta(z)$ is the Dirac delta-function, and we used the well-known representation of this function.

Then, taking $\varepsilon = 2\gamma\nu$ at $\gamma \to 0$ we obtain

$$\frac{\pi\hbar}{M\omega} Coth\left(\frac{\hbar\omega}{2k_B T}\right) = \frac{2\hbar}{M}\lim_{\gamma \to 0}\int_0^\infty d\nu\, Coth\left(\frac{\hbar\nu}{2k_B T}\right)\frac{2\gamma\nu}{(\nu^2 - \omega^2)^2 + 4\gamma^2\nu^2} = 2\pi\sigma^2(FDT), \tag{65}$$

where $\sigma^2(FDT)$ is in accordance with Eq,(60) but at the very restrictive condition $\gamma \to 0$.

Taking into account Eq.(65) we can to rewrite Eq.(63) as follows

$$\rho(x) = [2\pi\sigma^2(FDT)]^{-1/2} \exp\left[-x^2 / 2\sigma^2(FDT)\right], \tag{66}$$

formally coinciding with Eqs.(62).

Thus, we can to conclude that the Eq.(63) corresponds to the case of the infinitesimally weak coupling of the oscillator with a bath relating to the infinite time of relaxation. In other words, this is a case of the infinitesimally thin spectral line.

Contrary to the special case of Eq.(66) with $\gamma \to 0$, the formulas in Eq.(62) describe a general case including a finite value of $\gamma$.

Furthermore, Eqs.(54)-(56) allow to demonstrate a temporal behavior of the spatial variances and the covariance at any $t \in [0, \infty]$ and at different set of temperatures.

Figure 2 shows the temporal behavior of normalized variances $\sigma_{1,2}^2(t)/\sigma_{1,2}^2(FDT)$ for the first (curve 1) and second (curve 2) oscillators connected with different thermostats at equal temperatures $T_1 = T_2 = 300K$ in accordance with Eqs.(54), (55) and (60). We have chosen the following parameters of oscillators: $m_1 = 10^{-23} g$, $m_2 = 3m_1$, $\omega_{01} = 10^{13} rad/s$, $\omega_{02} = 2\omega_{01}$, $\gamma_1 = 0.01\omega_{01}$, $\gamma_2 = 0.01\omega_{02}$, relating to the typical characteristics for solid materials. In figure 2(a) the initial variances are $\sigma_{1,2}^2(t=0) = \hbar / 2m_{1,2}\omega_{01,02}$ relating to the natural case of the cold system at $T_1 = T_2 = 0K$ in Eq.(63) and in figure 2(b) the initial variances $\sigma_1^2(t=0) = \hbar / 2m_1\omega_{01}$ and $\sigma_2^2(t=0) = 10(\hbar / 2m_2\omega_{02})$ corresponding to the arbitrary state of the second oscillator.

Normalization was done to the variances of oscillators in equilibrium at $300K$ in accordance with the Eq.(60) and indicated as $\sigma_{1,2}^2(FDT)$. Initial ratios are $\sigma_1^2(0)/\sigma_1^2(FDT) = 0.13$ for the first and $\sigma_2^2(0)/\sigma_2^2(FDT) = 0.25$ for the second oscillators in figure 2(a), and $\sigma_1^2(0)/\sigma_1^2(FDT) = 0.13$ for the first and $\sigma_2^2(0)/\sigma_2^2(FDT) = 2.5$ for the second oscillators in figure 2(b).

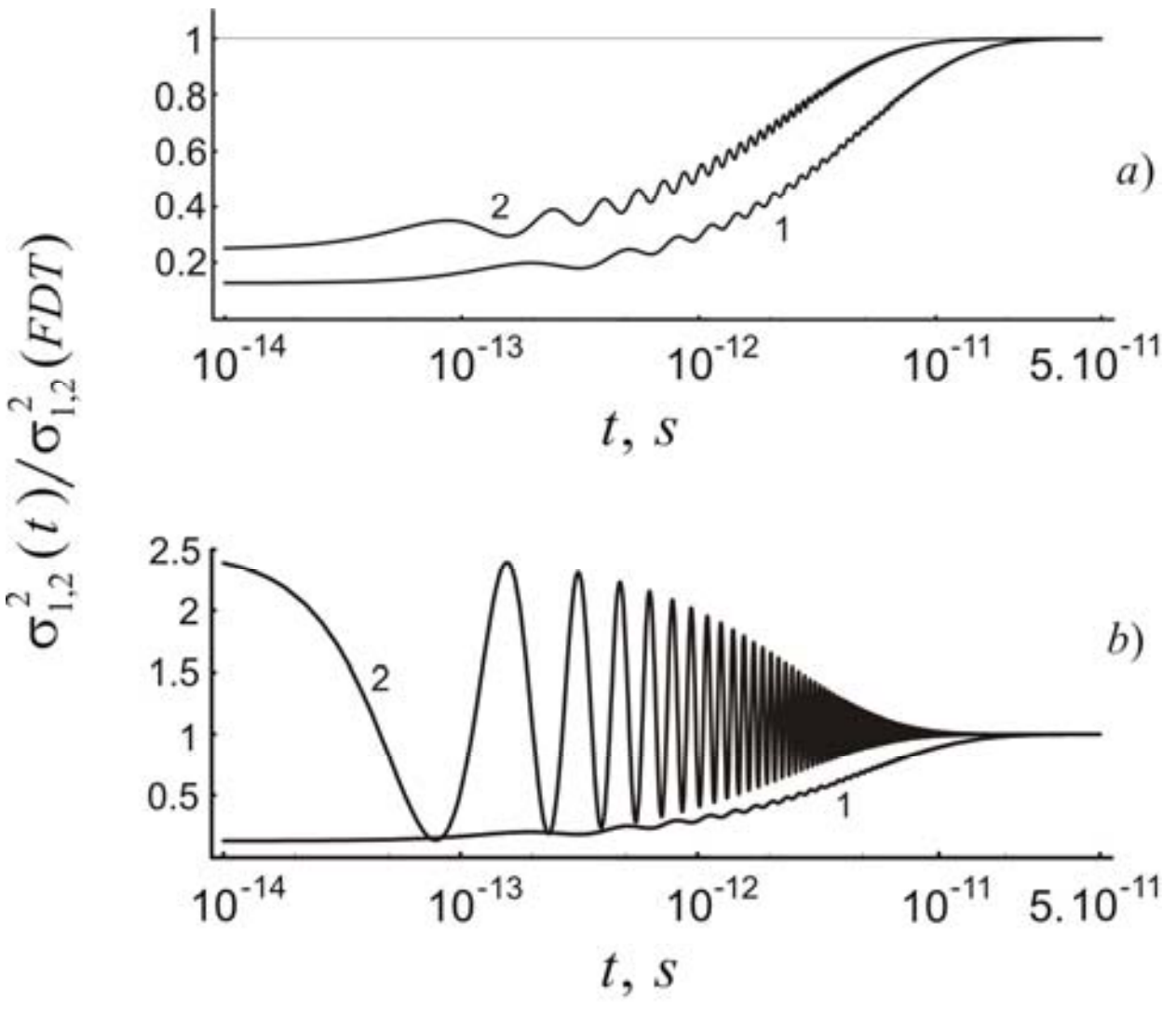

Fig.2

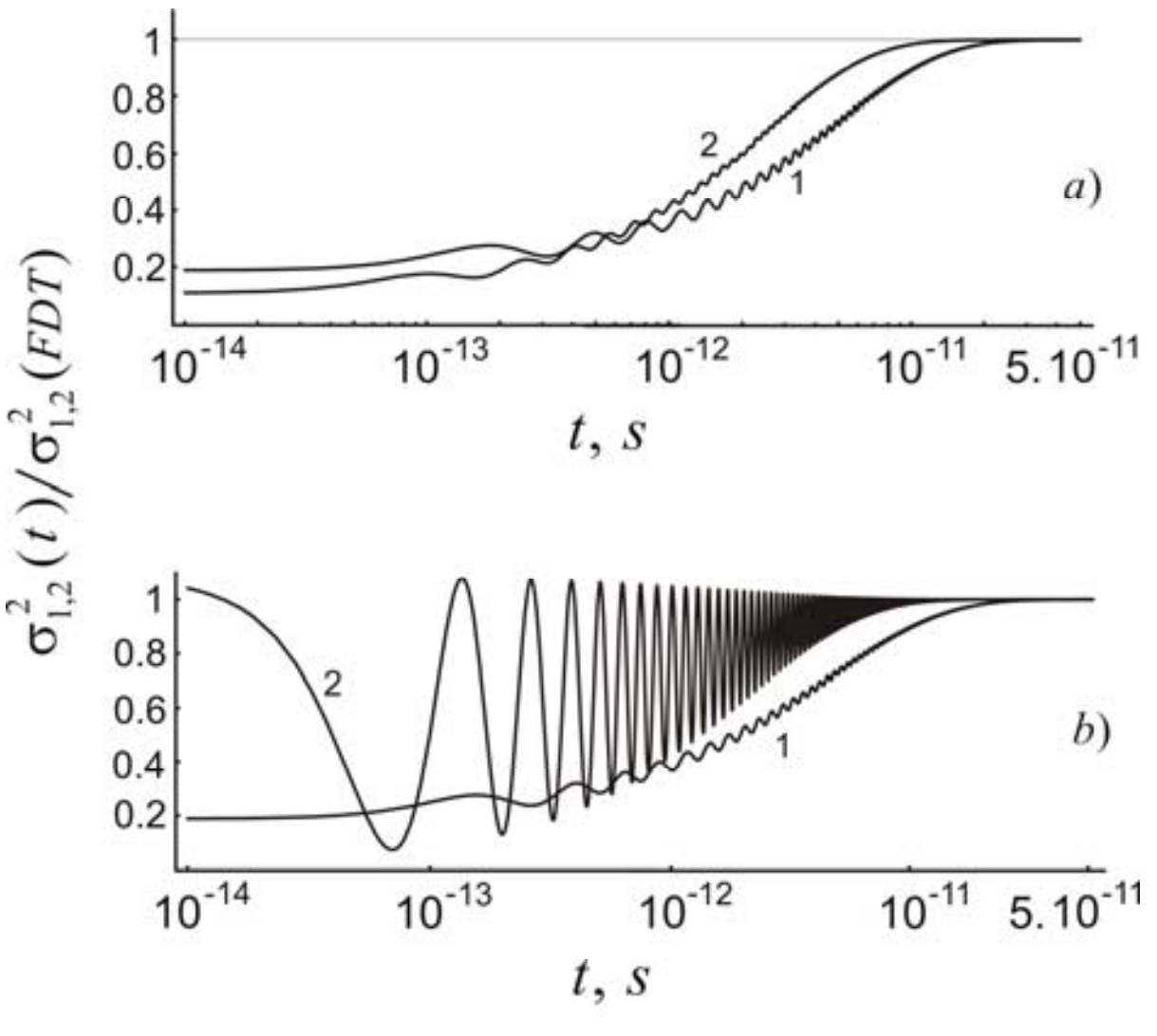

Fig.3

It is clearly seen that despite on the initial variances the final states (at $t \to \infty$) are always in accordance with the fluctuation dissipation theorem. Furthermore, the total temporal dynamics of each oscillator can be divided into three strongly marked phases: the early stage ($t < \Omega_{1,2}^{-1}$), the later transient stage ($\Omega_{1,2}^{-1} < t < \gamma_{1,2}^{-1}$), and the terminal time stage ($t > \gamma_{1,2}^{-1}$) of full equilibration of the each subsystem. Effect of difference in temperatures of the subsystems on the temporal dynamics is demonstrated in figure 3 as compare with figure 2 for two selected oscillators of the same parameters. In this figure the temporal behavior of normalized variances $\sigma_{1,2}^2(t)/\sigma_{1,2}^2(FDT)$ for the first 1 and second 2 oscillators connected with different thermostats at different temperatures $T_1 = 200K$ and $T_2 = 700K$ in accordance with Eqs.(54), (55) and (60) is shown. As well as in Fig.2 in figure 3(a) the initial variances are $\sigma_{1,2}^2(t=0) = \hbar / 2m_{1,2}\omega_{01,02}$ and in figure 3(b) the initial variances $\sigma_1^2(t=0) = \hbar / 2m_1\omega_{01}$ and $\sigma_2^2(t=0) = 10(\hbar / 2m_2\omega_{02})$. Normalization was done to the variances of oscillators in equilibrium at $200K$ and $700K$ correspondingly. Initial ratios in this case are $\sigma_1^2(0)/\sigma_1^2(FDT) = 0.19$ for the first, $\sigma_2^2(0)/\sigma_2^2(FDT) = 0.11$ for the second oscillators in figure 3(a), and $\sigma_1^2(0)/\sigma_1^2(FDT) = 0.19$ for the first and $\sigma_2^2(0)/\sigma_2^2(FDT) = 1.1$ for the second oscillators in figure 3(b). Despite on the other thermodynamic conditions the final states are always in accordance with the fluctuation dissipation theorem, too. Moreover, the total temporal dynamics of each oscillator can be divided into same three strongly marked phases up to full equilibration of the each subsystem.

The next figure exemplifies the pure quantum conditions ($T_1 = T_2 = 0K$) in the system under our study. The temporal dynamics of normalized variances $\sigma_{1,2}^2(t)/\sigma_{1,2}^2(FDT)$ for the first 1 and second

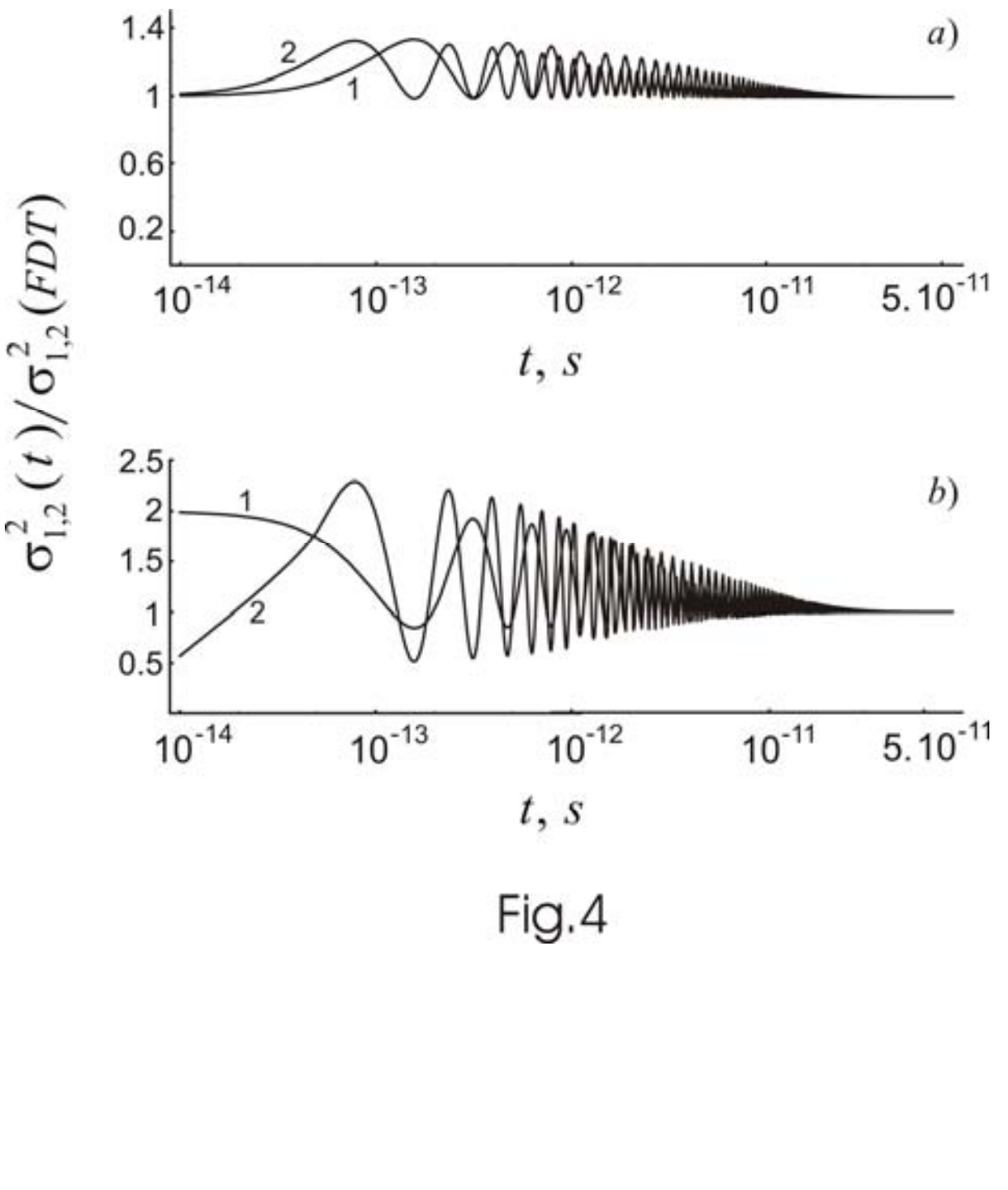

Fig.4

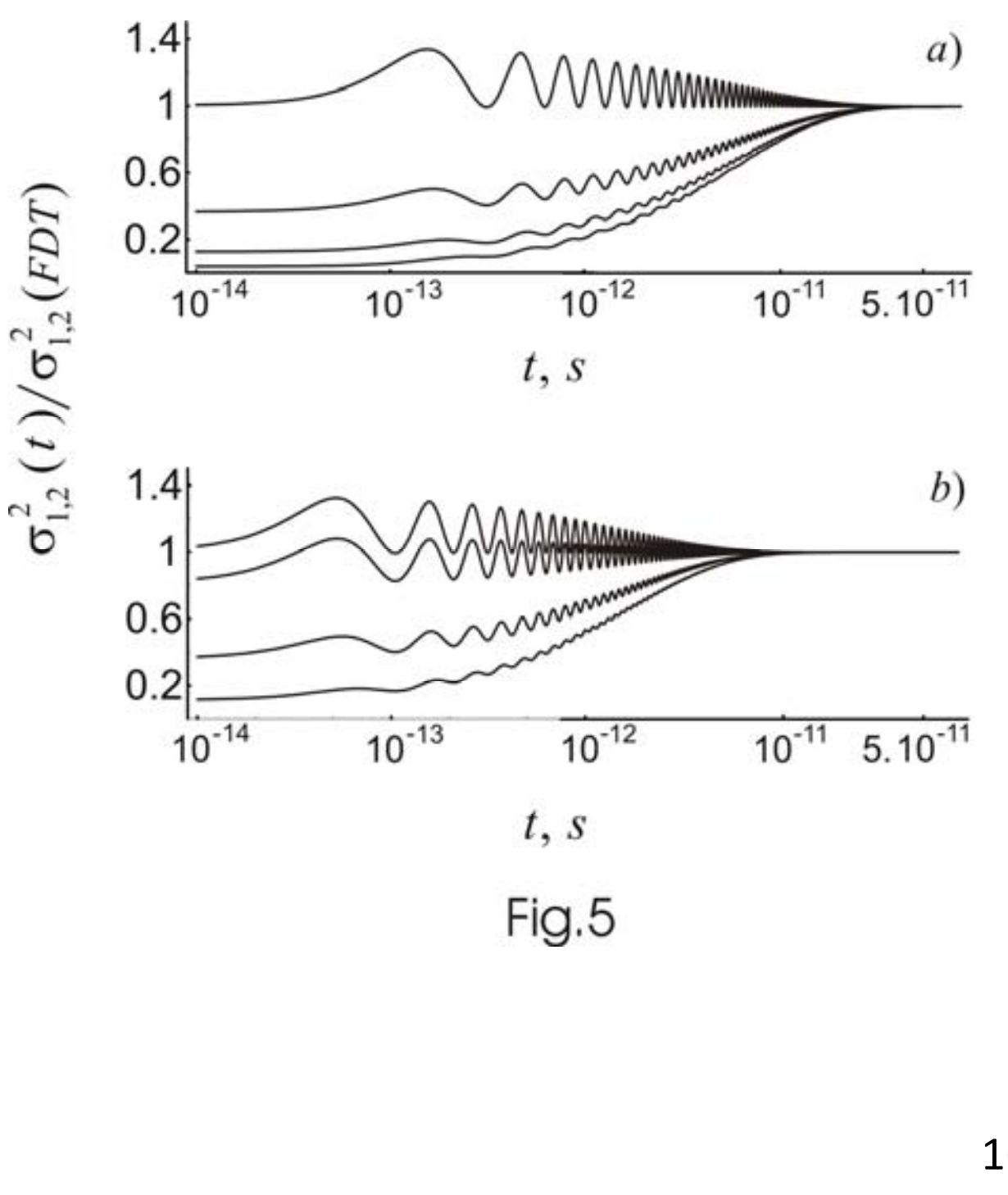

Fig.5

2 oscillators connected with different thermostats at absolute zero temperature $T_1 = T_2 = 0K$ in accordance with Eqs.(54), (55) and (60) is demonstrated in figure 4. In figure 4(a) the initial variances are $\sigma_{1,2}^2(t=0) = \hbar / 2m_{1,2}\omega_{01,02}$ corresponding to the subsystems at $T_1 = T_2 = 0K$ and normalization was done to the variances of oscillators in equilibrium at $T_1 = T_2 = 0K$ in accordance with the Eq.(60). It is seen that the initial ratio $\sigma_{1,2}^2(0)/\sigma_{1,2}^2(FDT) = 1$ will be the same after the time transient period $\Omega_{1,2}^{-1} < t < \gamma_{1,2}^{-1}$ due to connections with reservoirs. In figure 4(b) the initial variances $\sigma_1^2(t=0) = 2(\hbar / 2m_1\omega_{01})$ and $\sigma_2^2(t=0) = 0.5(\hbar / 2m_2\omega_{02})$ corresponding to the arbitrary states of these oscillators. Normalization was also done to the variances at absolute zero temperature, in accordance with the Eq.(60). This figure as well as the previous two figures indicates that the final states are always in accordance with the fluctuation dissipation theorem in spite of initial disturbances of any strength. In addition, to illustrate our conclusion in detail, figure 5 shows the normalized variances $\sigma_{1,2}^2(t)/\sigma_{1,2}^2(FDT)$ as the functions of time in accordance with Eqs.(54), (55) and (60) for the first oscillator in figure 5(a) and for the second oscillator in figure 5(b) connected with different thermostats at different temperatures $T = 10K$ , $T = 100K$ , $T = 300K$ and $T = 1000K$ (from the upper to low curve in each figures). Parameters of oscillators are $m_2 = 5m_1$, $\omega_{02} = 3\omega_{01}$ , $\gamma_{1,2} = 0.01\omega_{01,02}$, where $m_1 = 10^{-23}g$ , $\omega_{01} = 10^{13} rad/s$ . Initial variances are $\sigma_{1,2}^2(t=0) = \hbar / 2m_{1,2}\omega_{01,02}$ and normalization here was done to the variances in equilibrium at corresponding temperatures.

Another characteristics of the system is the covariance $\beta_{12}(t, \lambda, T_1, T_2)$ in Eq.(56). This quantity depends on the coupling constant and on both temperatures simultaneously. Figure 6 exemplifies the temporal dynamics of the normalized covariance $\beta_{12}(t)/\sqrt{\sigma_1^2\sigma_2^2(FDT)}$ for the coupled oscillators connected with different thermostats at different temperatures $T_1 = 200K$ and $T_2 = 700K$ in accordance with Eqs.(56),(60). Parameters of two coupled oscillators $m_1 = 10^{-23}g$, $m_2 = 3m_1$, $\omega_{01} = 10^{13} rad/s$, $\omega_{02} = 2\omega_{01}$, $\gamma_1 = 0.01\omega_{01}$, $\gamma_2 = 0.01\omega_{02}$ are the same for each figure. Coupling parameter $\rho$ is shown in inserts for each picture. Normalization was done to the variances of oscillators in equilibrium at $200K$ and $700K$ correspondingly, in accordance with the Eq.(60).

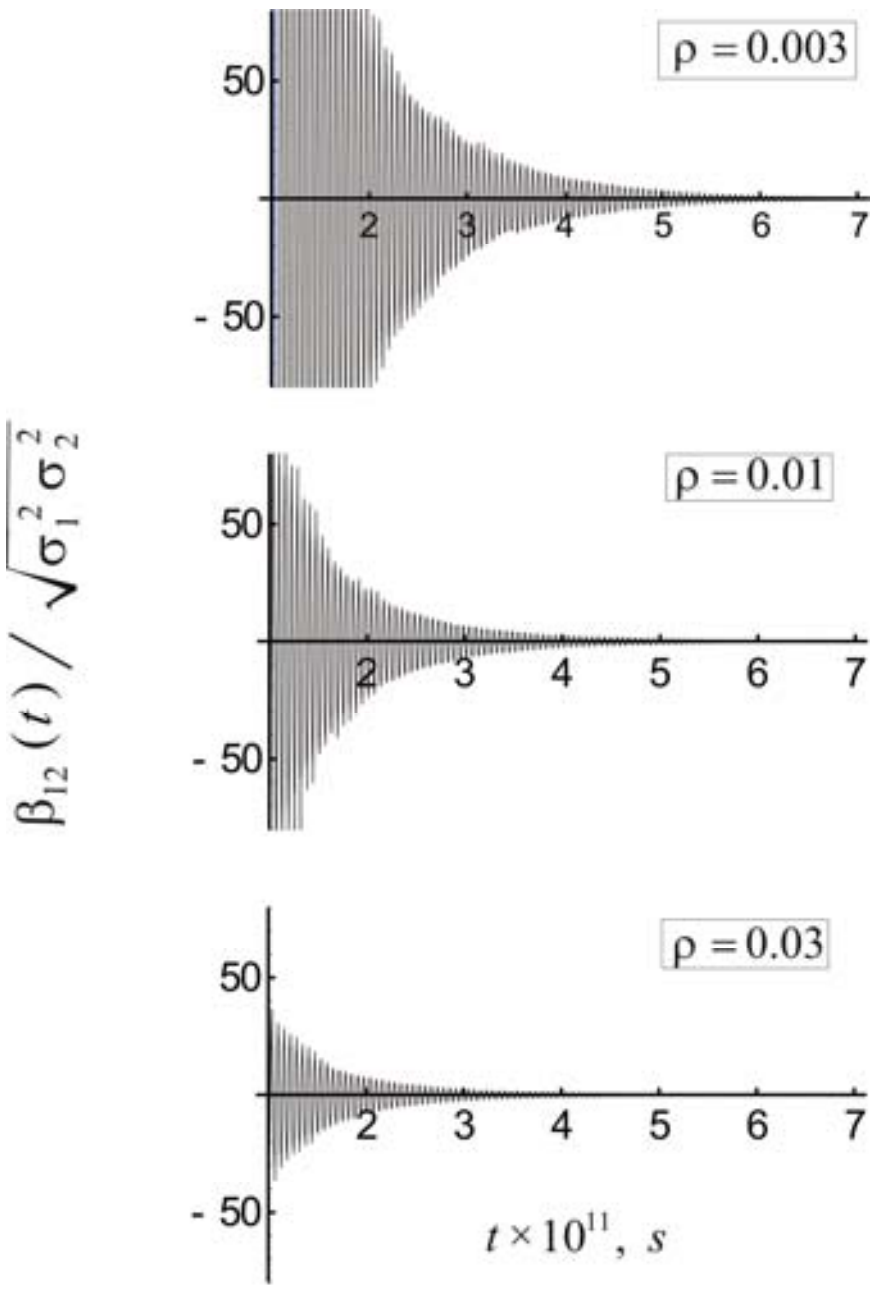

Fig.6

It should be noted that despite of the same $\gamma_{1,2}$ for each figure the rate of zeroing of the covariance is different due to different coupling parameter $\rho$ or coupling constant $\lambda$, see Eq.(34). The larger the $\rho$, the shorter is the time of zeroing. Because corresponding inverse matrix element is equal to $\beta_{12}^{-1} = \beta_{12}/(\beta_{12}^2 - \beta_{11}\beta_{22})$, the process $\beta_{12}(t) \to 0$ means simultaneously $\beta_{12}^{-1}(t) \to 0$. That is why in the long time regime there is good reason to believe that our subsystems in quasi-equilibrium at their own temperatures even despite of the arbitrary difference in temperatures within the whole system.

## IV. Conclusion

Our paper is devoted to the study of the relaxation problem of open quantum systems. Using the path integral methods basically described in [45-47] we found an analytical expression for time-dependent density matrix of two coupled quantum oscillators interacting with different reservoirs of oscillators. We calculated the Feynman-Vernon functional for this case and corresponding propagator in order to find the temporally dependent reduced density matrix for two coupled quantum oscillators. As for initial conditions we assumed a factorized form for the initial density matrix of two oscillators relating to the independent states of two initially "cold" particles. The analytical expression for the temporally dependent density matrix was found in the linear regime with respect to the coupling constant between the oscillators. In this regime the expression for the density matrix in Eqs.(53)-(57) takes the Gaussian form. From the general formula we can to obtain all known expressions for density matrix in steady-state regime. Time-dependent spatial variances and covariance were investigated analytically and numerically. In the short-time limit the variances and covariance are identical to their initially given values, as it must be. It was shown that asymptotic variances in the long-time limit are always in accordance with the fluctuation dissipation theorem despite on their initial values. The principal conclusion is that in the weak coupling approach there is good reason to believe that subsystems asymptotically in equilibrium at their own temperatures even despite of the arbitrary difference in temperatures within the whole system.

## Acknowledgement

Thank you for the reading of the paper.

## Appendix A. Solution to the motion equations in Eq.(30).

After exchanging the variables $X_{1,2} = x_{1,2} + y_{1,2}$ and $\xi_{1,2} = x_{1,2} - y_{1,2}$ in Eq.(30) we have two pairs of coupled equations

$$\begin{cases} \ddot{X}_1 + 2\gamma_1 \dot{X}_1 + \omega_{01}^2 X_1 = (\lambda / M_1) X_2 \\ \ddot{X}_2 + 2\gamma_2 \dot{X}_2 + \omega_{02}^2 X_2 = (\lambda / M_2) X_1 \end{cases}, \tag{A1}$$

$$\begin{cases} \ddot{\xi}_1 - 2\gamma_1 \dot{\xi}_1 + \omega_{01}^2 \xi_1 = (\lambda / M_1) \xi_2 \\ \ddot{\xi}_2 - 2\gamma_2 \dot{\xi}_2 + \omega_{02}^2 \xi_2 = (\lambda / M_2) \xi_1 \end{cases}, \tag{A2}$$

We will seek a solution, for example of the Eq. (A1) using the following form $X_{1,2} = A_{1,2} \sin(\omega\tau + \varphi) \exp(-\delta\tau)$. From the nontriviality condition one obtains a determinant equation, which yields

$$\omega^4 - (\omega_{01}^2 + \omega_{02}^2)\omega^2 + \omega_{01}^2 \omega_{02}^2 - \lambda^2 / M_1 M_2 = 0, \tag{A3}$$

the roots of the equation are

$$\Omega_{1,2}^2 = (\omega_{01}^2 + \omega_{02}^2)/2 \mp \sqrt{(\omega_{02}^2 - \omega_{01}^2)^2 / 4 + \lambda^2 / M_1 M_2}, \tag{A4}$$

$$\delta_{1,2} = \frac{(\Omega_{1,2}^2 - \omega_{02}^2)\gamma_1 + (\Omega_{1,2}^2 - \omega_{01}^2)\gamma_2}{(\Omega_{1,2}^2 - \omega_{02}^2) + (\Omega_{1,2}^2 - \omega_{01}^2)}, \tag{A5}$$

where we neglected in the determinant equation by the terms containing $\gamma_{1,2}, \delta$ to the second and larger exponents. One of the reasons to do so is the equation $\delta^2 - (\gamma_1 + \gamma_2)\delta + \gamma_1 \gamma_2 = 0$ , which follows from the same determinant equation.

To construct a general solution we follow the known receipt from, for example [50-52]

$$\tilde{X}_1(\tau) = A_1 \sin(\Omega_1\tau + \varphi_1)\exp(-\delta_1\tau) + r_2 A_2 \sin(\Omega_2\tau + \varphi_2)\exp(-\delta_2\tau),$$
$$\tilde{X}_2(\tau) = r_1 A_1 \sin(\Omega_1\tau + \varphi_1)\exp(-\delta_1\tau) + A_2 \sin(\Omega_2\tau + \varphi_2)\exp(-\delta_2\tau) \tag{A6}$$

where the coefficient $r_1$ determines a relative contribution to the first eigenmode from the second oscillator and vice versa for $r_2$. They are as follows

$$r_1 = \frac{\omega_{01}^2 - \Omega_1^2}{\lambda / M_1}, \quad r_2 = \frac{\omega_{02}^2 - \Omega_2^2}{\lambda / M_2}. \tag{A7}$$

Solutions of the system (A1) in the general form satisfying the conditions $X_{1,2}(0) = X_{i1,2}$ and $X_{1,2}(t) = X_{f1,2}$ are as follows

$$\begin{aligned}\tilde{X}_1(\tau) = {} & \left[\frac{X_{f1} - r_2 X_{f2}}{(1-r_1r_2)\sin(\Omega_1 t)}\exp(\delta_1 t) - \cot(\Omega_1 t)\frac{X_{i1} - r_2 X_{i2}}{(1-r_1r_2)}\right]\sin(\Omega_1\tau)\exp(-\delta_1\tau) + \\ & + \frac{X_{i1} - r_2 X_{i2}}{(1-r_1r_2)}\cos(\Omega_1\tau)\exp(-\delta_1\tau) + \\ & + \left[\frac{X_{f2} - r_1 X_{f1}}{(1-r_1r_2)\sin(\Omega_2 t)}\exp(\delta_2 t) - \cot(\Omega_2 t)\frac{X_{i2} - r_1 X_{i1}}{(1-r_1r_2)}\right] r_2 \sin(\Omega_2\tau)\exp(-\delta_2\tau) + \\ & + \frac{X_{i2} - r_1 X_{i1}}{(1-r_1r_2)} r_2 \cos(\Omega_2\tau)\exp(-\delta_2\tau)\end{aligned}, \tag{A8}$$

$$\begin{aligned}\tilde{X}_2(\tau) = {} & \left[\frac{X_{f1} - r_2 X_{f2}}{(1-r_1r_2)\sin(\Omega_1 t)}\exp(\delta_1 t) - \cot(\Omega_1 t)\frac{X_{i1} - r_2 X_{i2}}{(1-r_1r_2)}\right] r_1 \sin(\Omega_1\tau)\exp(-\delta_1\tau) + \\ & + \frac{X_{i1} - r_2 X_{i2}}{(1-r_1r_2)} r_1 \cos(\Omega_1\tau)\exp(-\delta_1\tau) + \\ & + \left[\frac{X_{f2} - r_1 X_{f1}}{(1-r_1r_2)\sin(\Omega_2 t)}\exp(\delta_2 t) - \cot(\Omega_2 t)\frac{X_{i2} - r_1 X_{i1}}{(1-r_1r_2)}\right]\sin(\Omega_2\tau)\exp(-\delta_2\tau) + \\ & + \frac{X_{i2} - r_1 X_{i1}}{(1-r_1r_2)}\cos(\Omega_2\tau)\exp(-\delta_2\tau)\end{aligned}. \tag{A9}$$

By the same way we obtained the solutions of the system (A2) satisfying the conditions $\xi_{1,2}(0) = \xi_{i1,2}$ and $\xi_{1,2}(t) = \xi_{f1,2}$

$$\begin{aligned}\tilde{\xi}_1(\tau) = {} & \left[\frac{\xi_{f1} - r_2 \xi_{f2}}{(1-r_1r_2)\sin(\Omega_1 t)}\exp(-\delta_1 t) - \cot(\Omega_1 t)\frac{\xi_{i1} - r_2 \xi_{i2}}{(1-r_1r_2)}\right]\sin(\Omega_1\tau)\exp(\delta_1\tau) + \\ & + \frac{\xi_{i1} - r_2 \xi_{i2}}{(1-r_1r_2)}\cos(\Omega_1\tau)\exp(\delta_1\tau) + \\ & + \left[\frac{\xi_{f2} - r_1 \xi_{f1}}{(1-r_1r_2)\sin(\Omega_2 t)}\exp(-\delta_2 t) - \cot(\Omega_2 t)\frac{\xi_{i2} - r_1 \xi_{i1}}{(1-r_1r_2)}\right] r_2 \sin(\Omega_2\tau)\exp(\delta_2\tau) + \\ & + \frac{\xi_{i2} - r_1 \xi_{i1}}{(1-r_1r_2)} r_2 \cos(\Omega_2\tau)\exp(\delta_2\tau)\end{aligned}, \tag{A10}$$

$$\tilde{\xi}_2(\tau)=\left[\frac{\xi_{f1}-r_2\xi_{f2}}{(1-r_1r_2)\sin(\Omega_1 t)}\exp(-\delta_1 t)-\cot(\Omega_1 t)\frac{\xi_{i1}-r_2\xi_{i2}}{(1-r_1r_2)}\right]r_1\sin(\Omega_1\tau)\exp(\delta_1\tau)+ \\ +\frac{\xi_{i1}-r_2\xi_{i2}}{(1-r_1r_2)}r_1\cos(\Omega_1\tau)\exp(\delta_1\tau)+ \\ +\left[\frac{\xi_{f2}-r_1\xi_{f1}}{(1-r_1r_2)\sin(\Omega_2 t)}\exp(-\delta_2 t)-\cot(\Omega_2 t)\frac{\xi_{i2}-r_1\xi_{i1}}{(1-r_1r_2)}\right]\sin(\Omega_2\tau)\exp(\delta_2\tau)+ \\ +\frac{\xi_{i2}-r_1\xi_{i1}}{(1-r_1r_2)}\cos(\Omega_2\tau)\exp(\delta_2\tau) \quad . \qquad \text{(A11)}$$

It is clearly seen from Eqs.(A8)-(A11) and Eqs.(A4),(A7), that in case of uncoupled oscillators ( $\lambda=0$ ) we have two independent trajectories identically coinciding with Eqs.(6.10),(6.11) from [45].

## Appendix B. Coefficients in Eqs.(38),(39).

Taking into account Eqs. (A8)-(A11) the part of action in Eq.(37) in new variables is expressed as follows

$$\tilde{S}_{cl}^{(1)}+\tilde{S}_{cl}^{(2)}=\int_0^t dt'\{M_1\dot{\tilde{X}}_1\dot{\tilde{\xi}}_1/2-M_1\omega_1^2\tilde{X}_1\tilde{\xi}_1/2-M_1\gamma_1\dot{\tilde{X}}_1\tilde{\xi}_1 \\ +M_2\dot{\tilde{X}}_2\dot{\tilde{\xi}}_2/2-M_2\omega_2^2\tilde{X}_2\tilde{\xi}_2/2-M_2\gamma_2\dot{\tilde{X}}_2\tilde{\xi}_2\}, \qquad \text{(B1)}$$

where after an integration we have

$$\tilde{S}_{cl}^{(1)}+\tilde{S}_{cl}^{(2)}=D_1(X_{f1}\xi_{f1})+[D_5+D_5'](X_{f2}\xi_{f1})+[D_6+D_6'](X_{f1}\xi_{f2})+D_1'(X_{f2}\xi_{f2})+ \\ +D_2(X_{i1}\xi_{f1})+D_3(X_{f1}\xi_{i1})+D_4(X_{i1}\xi_{i1})+D_2'(X_{i2}\xi_{f2})+D_3'(X_{f2}\xi_{i2})+D_4'(X_{i2}\xi_{i2})+ \\ +[D_7+D_7'](X_{i1}\xi_{f2})+[D_8+D_8'](X_{i2}\xi_{f1})+[D_9+D_9'](X_{f1}\xi_{i2})+[D_{10}+D_{10}'](X_{f2}\xi_{i1})+ \\ +[D_{11}+D_{11}'](X_{i1}\xi_{i2})+[D_{12}+D_{12}'](X_{i2}\xi_{i1}) \quad , \qquad \text{(B2)}$$

where

$$D_1(t)=(M_1/2)n_1\bar{n}_1b_1,\ D_2(t)=(M_1/2)(\bar{n}_1b_2-m_1\bar{n}_1b_1),\ D_3(t)=(M_1/2)(n_1b_3-m_1n_1b_1), \\ D_4(t)=(M_1/2)(m_1^2b_2-m_1b_2-m_1b_3+b_4),\ D_5(t)=r_2(M_1/2)(n_2\bar{n}_1b_5-n_1\bar{n}_1b_1), \\ D_6(t)=r_2(M_1/2)(n_1\bar{n}_2b_9-n_1\bar{n}_1b_1),\ D_7(t)=r_2(M_1/2)(m_1\bar{n}_1b_1-\bar{n}_1b_2-m_1\bar{n}_2b_9+\bar{n}_2b_{10}), \\ D_8(t)=r_2(M_1/2)(m_1\bar{n}_1b_1-\bar{n}_1b_2-m_2\bar{n}_1b_5+\bar{n}_1b_6), \\ D_9(t)=r_2(M_1/2)(n_1m_1b_1-n_1b_3-n_1m_2b_9+n_1b_{11}), \\ D_{10}(t)=r_2(M_1/2)(n_1m_1b_1-n_1b_3-n_2m_1b_5+n_2b_7), \\ D_{11}(t)=r_2(M_1/2)(m_1b_2-m_1^2b_1+m_1b_3-b_4+m_1m_2b_9-m_2b_{10}-m_1b_{11}+b_{12}), \\ D_{12}(t)=r_2(M_1/2)(m_1b_2-m_1^2b_1+m_1b_3-b_4+m_1m_2b_5-m_1b_6-m_2b_7+b_8), \quad , \qquad \text{(B3)}$$

$$
\begin{aligned}
&D_1'(t) = (M_2/2)n_2\bar{n}_2 b_1',\ D_2'(t) = (M_2/2)(\bar{n}_2 b_2' - m_2\bar{n}_2 b_1'),\ D_3'(t) = (M_2/2)(n_2 b_3' - m_2 n_2 b_1'),\\
&D_4'(t) = (M_2/2)(m_2^2 b_2' - m_2 b_2' - m_2 b_3' + b_4'),\ D_5'(t) = r_1(M_2/2)(n_2\bar{n}_1 b_5' - n_2\bar{n}_2 b_1'),\\
&D_6'(t) = r_1(M_2/2)(n_1\bar{n}_2 b_9' - n_2\bar{n}_2 b_1'),\ D_7'(t) = r_1(M_2/2)(m_2\bar{n}_2 b_1' - n_2 b_2' - m_1\bar{n}_2 b_9' + \bar{n}_2 b_{10}'),\\
&D_8'(t) = r_1(M_2/2)(m_2\bar{n}_2 b_1' - n_2 b_2' - m_2 n_1 b_5' + \bar{n}_1 b_6'),\\
&D_9'(t) = r_1(M_2/2)(n_2 m_2 b_1' - n_2 b_3' - n_1 m_2 b_9' + n_1 b_{11}'),\\
&D_{10}'(t) = r_1(M_2/2)(n_2 m_2 b_1' - n_2 b_3' - n_2 m_1 b_5' + n_2 b_8'),\\
&D_{11}'(t) = r_1(M_2/2)(m_2 b_2' - m_2^2 b_1' + m_2 b_3' - b_4' + m_1 m_2 b_9' - m_2 b_{10}' - m_1 b_{11}' + b_{12}'),\\
&D_{12}'(t) = r_1(M_2/2)(m_2 b_2' - m_2^2 b_1' + m_2 b_3' - b_4' + m_1 m_2 b_5' - m_1 b_6' - m_2 b_8' + b_7'),
\end{aligned}
\quad , \quad \text{(B4)}
$$

where $n_{1,2}(t) = \exp(\delta_{1,2}t)/\sin(\Omega_{1,2}t)$, $\bar{n}_{1,2}(t) = \exp(-\delta_{1,2}t)/\sin(\Omega_{1,2}t)$, $m_{1,2}(t) = \cot(\Omega_{1,2}t)$,

$$
\begin{aligned}
&b_1 = \Omega_1^2 s_1 - \omega_{01}^2 s_2 - 2\Omega_1\gamma_1 s_5,\\
&b_2 = -\Omega_1^2 s_5 - \omega_{01}^2 s_5 - \Omega_1\delta_1(s_1 + s_2) + 2\Omega_1\gamma_1 s_2,\\
&b_3 = -\Omega_1^2 s_5 - \omega_{01}^2 s_5 + \Omega_1\delta_1(s_1 + s_2) - 2\Omega_1\gamma_1 s_1,\\
&b_4 = \Omega_1^2 s_2 - \omega_{01}^2 s_1 + 2\Omega_1\gamma_1 s_5,\\
&b_5 = r_2(\Omega_1\Omega_2 s_7 - \omega_{01}^2 s_{10} - \Omega_1\delta_2 s_8 + \Omega_2\delta_1 s_9 - 2\Omega_2\gamma_1 s_9),\\
&b_6 = r_2(-\Omega_1\Omega_2 s_8 - \omega_{01}^2 s_9 - \Omega_1\delta_2 s_7 - \Omega_2\delta_1 s_{10} + 2\Omega_2\gamma_1 s_{10}),\\
&b_7 = r_2(-\Omega_1\Omega_2 s_9 - \omega_{01}^2 s_8 + \Omega_1\delta_2 s_{10} + \Omega_2\delta_1 s_7 - 2\Omega_2\gamma_1 s_7),\\
&b_8 = r_2(\Omega_1\Omega_2 s_{10} - \omega_{01}^2 s_7 + \Omega_1\delta_2 s_9 - \Omega_2\delta_1 s_8 + 2\Omega_2\gamma_1 s_8),\\
&b_9 = r_2(\Omega_1\Omega_2 s_{11} - \omega_{01}^2 s_{14} - \Omega_2\delta_1 s_{12} + \Omega_1\delta_2 s_{13} - 2\Omega_1\gamma_1 s_{13}),\\
&b_{10} = r_2(-\Omega_1\Omega_2 s_{12} - \omega_{01}^2 s_{13} - \Omega_2\delta_1 s_{11} - \Omega_1\delta_2 s_{14} + 2\Omega_1\gamma_1 s_{14}),\\
&b_{11} = r_2(-\Omega_1\Omega_2 s_{13} - \omega_{01}^2 s_{12} + \Omega_2\delta_1 s_{14} + \Omega_1\delta_2 s_{11} - 2\Omega_1\gamma_1 s_{11}),\\
&b_{12} = r_2(\Omega_1\Omega_2 s_{14} - \omega_{01}^2 s_{11} + \Omega_2\delta_1 s_{13} - \Omega_1\delta_2 s_{12} + 2\Omega_1\gamma_1 s_{12}),
\end{aligned}
\quad \text{(B5)}
$$

$$
\begin{aligned}
b'_1 &= \Omega_2^2 s_3 - \omega_{02}^2 s_4 - 2\Omega_2\gamma_2 s_6, \\
b'_2 &= -\Omega_2^2 s_6 - \omega_{02}^2 s_6 - \Omega_2\delta_2(s_3+s_4) + 2\Omega_2\gamma_2 s_4, \\
b'_3 &= -\Omega_2^2 s_6 - \omega_{02}^2 s_6 + \Omega_2\delta_2(s_3+s_4) - 2\Omega_2\gamma_2 s_3, \\
b'_4 &= \Omega_2^2 s_4 - \omega_{02}^2 s_3 + 2\Omega_2\gamma_2 s_6, \\
b'_5 &= r_1(\Omega_1\Omega_2 s_7 - \omega_{02}^2 s_{10} - \Omega_1\delta_2 s_8 + \Omega_2\delta_1 s_9 - 2\Omega_2\gamma_2 s_9), \\
b'_6 &= r_1(-\Omega_1\Omega_2 s_8 - \omega_{02}^2 s_9 - \Omega_1\delta_2 s_7 - \Omega_2\delta_1 s_{10} + 2\Omega_2\gamma_2 s_{10}), \\
b'_7 &= r_1(\Omega_1\Omega_2 s_{10} - \omega_{02}^2 s_7 + \Omega_1\delta_2 s_9 - \Omega_2\delta_1 s_8 + 2\Omega_2\gamma_2 s_8), \\
b'_8 &= r_1(-\Omega_1\Omega_2 s_9 - \omega_{02}^2 s_8 + \Omega_1\delta_2 s_{10} + \Omega_2\delta_1 s_7 - 2\Omega_2\gamma_2 s_7), \\
b'_9 &= r_1(\Omega_1\Omega_2 s_{11} - \omega_{02}^2 s_{14} - \Omega_2\delta_1 s_{12} + \Omega_1\delta_2 s_{13} - 2\Omega_1\gamma_2 s_{13}), \\
b'_{10} &= r_1(-\Omega_1\Omega_2 s_{12} - \omega_{02}^2 s_{13} - \Omega_2\delta_1 s_{11} - \Omega_1\delta_2 s_{14} + 2\Omega_1\gamma_2 s_{14}), \\
b'_{11} &= r_1(-\Omega_1\Omega_2 s_{13} - \omega_{02}^2 s_{12} + \Omega_2\delta_1 s_{14} + \Omega_1\delta_2 s_{11} - 2\Omega_1\gamma_2 s_{11}), \\
b'_{12} &= r_2(\Omega_1\Omega_2 s_{14} - \omega_{02}^2 s_{11} + \Omega_2\delta_1 s_{13} - \Omega_1\delta_2 s_{12} + 2\Omega_1\gamma_2 s_{12}),
\end{aligned}
\tag{B6}
$$

The part of action $\tilde{S}_{cl}^{(12)}$ corresponding to an interaction of two oscillators in Eq. (37) is determined as follows

$$
\tilde{S}_{cl}^{(12)} = \int_0^t dt' (\lambda/2)[\tilde{X}_1(t')\tilde{\xi}_2(t') + \tilde{X}_2(t')\tilde{\xi}_1(t')], \tag{B7}
$$

where $\tilde{X}_{1,2}$ and $\tilde{\xi}_{1,2}$ are also given by Eqs.(A8)-(A11).

After substitution of these equations in to Eq. (B7) we obtain

$$
\begin{aligned}
\tilde{S}_{cl}^{(12)} &= \Pi_1(X_{f1}\xi_{f1}) + \Pi_2(X_{f1}\xi_{f2}) + \Pi_3(X_{f2}\xi_{f1}) + \Pi_4(X_{f2}\xi_{f2}) + \\
&+ \Pi_5(X_{f1}\xi_{i1}) + \Pi_6(X_{f1}\xi_{i2}) + \Pi_7(X_{f2}\xi_{i1}) + \Pi_8(X_{f2}\xi_{i2}) + \\
&+ \Pi_9(X_{i1}\xi_{f1}) + \Pi_{10}(X_{i1}\xi_{f2}) + \Pi_{11}(X_{i2}\xi_{f1}) + \Pi_{12}(X_{i2}\xi_{f2}) + \\
&+ \Pi_{13}(X_{i1}\xi_{i1}) + \Pi_{14}(X_{i1}\xi_{i2}) + \Pi_{15}(X_{i2}\xi_{i1}) + \Pi_{16}(X_{i2}\xi_{i2})
\end{aligned}
, \tag{B8}
$$

where

$$
\begin{aligned}
&\Pi_1(t) = r_1(\lambda/2)(2n_1\bar{n}_1 s_2 - n_1\bar{n}_2 s_{14} - n_2\bar{n}_1 s_{10}),\ \Pi_2(t) = (\lambda/2) n_1\bar{n}_2 s_{14},\\
&\Pi_3(t) = (\lambda/2) n_2\bar{n}_1 s_{10},\ \Pi_4(t) = r_2(\lambda/2)(2n_2\bar{n}_2 s_4 - \bar{n}_1 n_2 s_{10} - n_1\bar{n}_2 s_{14}),\\
&\Pi_5(t) = r_1(\lambda/2)(2n_1 s_5 - 2n_1 m_1 s_2 - n_2 s_8 + n_2 m_1 s_{10} - n_1 s_{12} + n_1 m_2 s_{14}),\\
&\Pi_6(t) = (\lambda/2)(n_1 s_{12} - n_1 m_2 s_{14}),\ \Pi_7(t) = (\lambda/2)(n_2 s_8 - n_2 m_1 s_{10}),\\
&\Pi_8(t) = r_2(\lambda/2)(2n_2 s_6 - 2n_2 m_2 s_4 - n_2 s_8 + m_1 m_2 s_{10} - n_1 s_{12} + n_1 m_2 s_{14}),\\
&\Pi_9(t) = r_1(\lambda/2)(2n_1 s_5 - 2m_1\bar{n}_1 s_2 - \bar{n}_1 s_9 + m_2\bar{n}_1 s_{10} - n_2 s_{13} + m_1\bar{n}_2 s_{14}),\\
&\Pi_{10}(t) = (\lambda/2)(\bar{n}_2 s_{13} - m_1\bar{n}_2 s_{14}),\ \Pi_{11}(t) = (\lambda/2)(\bar{n}_1 s_9 - \bar{n}_1 m_2 s_{10}),\\
&\Pi_{12}(t) = r_2(\lambda/2)(2\bar{n}_2 s_6 - 2\bar{n}_2 m_2 s_4 + m_2\bar{n}_1 s_{10} - \bar{n}_2 s_{13} + m_1\bar{n}_2 s_{14} - n_1 s_9),\\
&\Pi_{13}(t) = r_1(\lambda/2)(-s_7 + 2m_1^2 s_2 - 4m_1 s_5 + 2s_1 + m_2 s_8 + m_1 s_9 - m_1 m_2 s_{10} -\\
&-s_{11} + m_1 s_{12} + m_2 s_{13} - m_1 m_2 s_{14}),\\
&\Pi_{14}(t) = (\lambda/2)(m_1 m_2 s_{14} - m_2 s_{13} - m_1 s_{12} + s_{11}),\\
&\Pi_{15}(t) = (\lambda/2)(s_7 - m_2 s_8 - m_1 s_9 + m_1 m_2 s_{10}),\\
&\Pi_{16}(t) = r_2(\lambda/2)(m_2 s_{13} + m_1 s_{12} - s_{11} - m_1 m_2 s_{10} + m_1 s_9 + m_2 s_8 - s_7 +\\
&+2m_2^2 s_4 - 4m_2 s_6 + 2s_3 - m_1 m_2 s_{14}),
\end{aligned} \tag{B9}
$$

where

$s_1(t) = t/2 + \sin(2\Omega_1 t)/4\Omega_1,\ s_2(t) = t/2 - \sin(2\Omega_1 t)/4\Omega_1,$

$s_3(t) = t/2 + \sin(2\Omega_2 t)/4\Omega_2,\ s_4(t) = t/2 - \sin(2\Omega_2 t)/4\Omega_2,$

$s_5(t) = \sin^2(\Omega_1 t)/2\Omega_1,\ s_6(t) = \sin^2(\Omega_2 t)/2\Omega_2,$

$$
\begin{aligned}
s_7(t) = \frac{1}{2}\Bigg[\frac{-\delta_-}{\delta_-^2+\Omega_-^2} + \frac{-\delta_-}{\delta_-^2+\Omega_+^2} + \exp(t\delta_-)\Bigg(\frac{\delta_- cos(t\Omega_-) + \Omega_- sin(t\Omega_-)}{\delta_-^2+\Omega_-^2} +\\
+\frac{\delta_- cos(t\Omega_+) + \Omega_+ sin(t\Omega_+)}{\delta_-^2+\Omega_+^2}\Bigg)\Bigg],
\end{aligned} \tag{B10}
$$

$$
\begin{aligned}
s_8(t) = \frac{1}{2}\Bigg[\frac{-\Omega_-}{\delta_-^2+\Omega_-^2} + \frac{\Omega_+}{\delta_-^2+\Omega_+^2} + \exp(t\delta_-)\Bigg(\frac{\Omega_- cos(t\Omega_-) - \delta_- sin(t\Omega_-)}{\delta_-^2+\Omega_-^2} +\\
+\frac{-\Omega_+ cos(t\Omega_+) + \delta_- sin(t\Omega_+)}{\delta_-^2+\Omega_+^2}\Bigg)\Bigg],
\end{aligned} \tag{B11}
$$

$$
\begin{aligned}
s_9(t) = \frac{1}{2}\Bigg[\frac{\Omega_-}{\delta_-^2+\Omega_-^2} + \frac{\Omega_+}{\delta_-^2+\Omega_+^2} + \exp(t\delta_-)\Bigg(\frac{-\Omega_- cos(t\Omega_-) + \delta_- sin(t\Omega_-)}{\delta_-^2+\Omega_-^2} +\\
+\frac{-\Omega_+ cos(t\Omega_+) + \delta_- sin(t\Omega_+)}{\delta_-^2+\Omega_+^2}\Bigg)\Bigg],
\end{aligned} \tag{B12}
$$

$$s_{10}(t)=\frac{1}{2}\left[\frac{-\delta_-}{\delta_-^2+\Omega_-^2}+\frac{\delta_-}{\delta_-^2+\Omega_+^2}+\exp(t\delta_-)\left(\frac{\delta_- cos(t\Omega_-)+\Omega_- sin(t\Omega_-)}{\delta_-^2+\Omega_-^2}-\right.\right.$$
$$\left.\left.-\frac{\delta_- cos(t\Omega_+)+\Omega_+ sin(t\Omega_+)}{\delta_-^2+\Omega_+^2}\right)\right], \qquad \text{(B13)}$$

$$s_{11}(t)=\frac{1}{2}\left[\frac{\delta_-}{\delta_-^2+\Omega_-^2}+\frac{\delta_-}{\delta_-^2+\Omega_+^2}+\exp(-t\delta_-)\left(\frac{-\delta_- cos(t\Omega_-)+\Omega_- sin(t\Omega_-)}{\delta_-^2+\Omega_-^2}+\right.\right.$$
$$\left.\left.+\frac{-\delta_- cos(t\Omega_+)+\Omega_+ sin(t\Omega_+)}{\delta_-^2+\Omega_+^2}\right)\right], \qquad \text{(B14)}$$

$$s_{12}(t)=\frac{1}{2}\left[\frac{\Omega_-}{\delta_-^2+\Omega_-^2}+\frac{\Omega_+}{\delta_-^2+\Omega_+^2}+\exp(-t\delta_-)\left(\frac{-\Omega_- cos(t\Omega_-)-\delta_- sin(t\Omega_-)}{\delta_-^2+\Omega_-^2}-\right.\right.$$
$$\left.\left.-\frac{\Omega_+ cos(t\Omega_+)+\delta_- sin(t\Omega_+)}{\delta_-^2+\Omega_+^2}\right)\right], \qquad \text{(B15)}$$

$$s_{13}(t)=\frac{1}{2}\left[\frac{-\Omega_-}{\delta_-^2+\Omega_-^2}+\frac{\Omega_+}{\delta_-^2+\Omega_+^2}+\exp(-t\delta_-)\left(\frac{\Omega_- cos(t\Omega_-)+\delta_- sin(t\Omega_-)}{\delta_-^2+\Omega_-^2}-\right.\right.$$
$$\left.\left.-\frac{\Omega_+ cos(t\Omega_+)+\delta_- sin(t\Omega_+)}{\delta_-^2+\Omega_+^2}\right)\right], \qquad \text{(B16)}$$

$$s_{11}(t)=\frac{1}{2}\left[\frac{\delta_-}{\delta_-^2+\Omega_-^2}+\frac{-\delta_-}{\delta_-^2+\Omega_+^2}+\exp(-t\delta_-)\left(\frac{-\delta_- cos(t\Omega_-)+\Omega_- sin(t\Omega_-)}{\delta_-^2+\Omega_-^2}+\right.\right.$$
$$\left.\left.+\frac{\delta_- cos(t\Omega_+)-\Omega_+ sin(t\Omega_+)}{\delta_-^2+\Omega_+^2}\right)\right], \qquad \text{(B17)}$$

where $\delta_- = \delta_1 - \delta_2$, $\Omega_\pm = \Omega_1 \pm \Omega_2$ .

## Appendix C. Evaluation of the fluctuational integral $G$ in Eq.(47).

The evaluation is realized as well as in [45, 46] by use of an orthogonal set of sine functions

$$X'_{1,2} = C\sum_{n=1}^{\infty} X_n^{(1,2)} Sin(\omega_n \tau), \qquad \xi'_{1,2} = C\sum_{m=1}^{\infty} \xi_m^{(1,2)} Sin(\omega_m \tau), \qquad \text{(C1)}$$

where $\omega_n = n\pi / t$ , and $C$ is the normalized constant for further convenience.

Substitution of Eq.(C1) into Eq.(32) and using Eqs.(47),(48) yields

$$G=\lim_{N\to\infty} J^4\int \mathcal{D}X_{1n}\mathcal{D}X_{2n}\mathcal{D}\xi_{1n}\mathcal{D}\xi_{2n}\exp\frac{i}{\hbar}\left\{\frac{M_1tC^2}{4}\sum_{n=1}^{N}\left[\left(\frac{n\pi}{t}\right)^2-\omega_{01}^2+\frac{\lambda}{M_1}\right]\right\}X_n^{(1)}\xi_n^{(1)}$$
$$\times\exp\frac{i}{\hbar}\left\{\frac{M_2tC^2}{4}\sum_{n=1}^{N}\left[\left(\frac{n\pi}{t}\right)^2-\omega_{02}^2+\frac{\lambda}{M_2}\right]\right\}X_n^{(2)}\xi_n^{(2)}\exp\left(\sum_{n=1}^{N}f_1\xi_{1n}+g_1\xi_{1n}^2\right)\exp\left(\sum_{n=1}^{N}f_2\xi_{2n}+g_2\xi_{2n}^2\right), \quad (C2)$$

where $J$ is the Jacobean determinant, the functions $f_{1,2}, g_{1,2}$ are not dependent on the integration variables, their view is clear from Eqs.(47),(48).

The functional measure in Eq.(C2) is defined as usual with the normalization factor $A_{1,2}=(2\pi i\hbar\varepsilon/M_{1,2})^{1/2}$ [46], where $\varepsilon=t/N$. It follows from Eq.(C2) that the fluctuational integral can be represented as $G=G_1G_2$. In its turn, each of $G_1$, $G_2$ is a product of $N$ Riemann integrals like

$$G_1=\lim_{N\to\infty}\frac{J^2}{A^2}\int_{-\infty}^{\infty}\frac{dX_1^{(1)}}{A}\dots\int_{-\infty}^{\infty}\frac{dX_N^{(1)}}{A}\int_{-\infty}^{\infty}\frac{d\xi_1^{(1)}}{A}\dots\int_{-\infty}^{\infty}\frac{d\xi_N^{(1)}}{A}\exp\frac{i}{\hbar}\left\{\frac{M_1tC^2}{4}\sum_{n=1}^{N}\left[\left(\frac{n\pi}{t}\right)^2-\omega_{01}^2+\frac{\lambda}{M_1}\right]\right\}X_n^{(1)}\xi_n^{(1)}$$
$$\times\exp\left(\sum_{n=1}^{N}f_1\xi_{1n}+g_1\xi_{1n}^2\right), \quad (C3)$$

and corresponding expression for $G_2$.

One of these integrals is equal

$$\int_{-\infty}^{\infty}\frac{dX_n^{(1)}}{A}\exp\frac{i}{\hbar}\frac{M_1tC^2}{4}\left[\left(\frac{n\pi}{t}\right)^2-\omega_{01}^2+\frac{\lambda}{M_1}\right]X_n^{(1)}\xi_n^{(1)}=$$
$$\frac{2\pi}{A}\delta[\xi_n^{(1)}]\left|\frac{M_1tC^2}{4\hbar}\left[\left(\frac{n\pi}{t}\right)^2-\omega_{01}^2+\frac{\lambda}{M_1}\right]\right|^{-1}, \quad (C4)$$

where $\delta[\xi_n^{(1)}]$ is the Dirac delta-function.

Then, a trivial integration with the Dirac delta-function finally yields

$$G_i=\lim_{N\to\infty}\frac{J^2}{A_i^2}\left(\frac{2\pi}{A_i^2}\right)^N\left(\frac{4\hbar}{M_itC^2}\right)^N\prod_{n=1}^{N}\left|\left[\omega_n^2-\omega_{0i}^2+\lambda/M_i\right]\right|^{-1}, \qquad (i=1,2). \quad (C5)$$

where from we obtain

$$G_i=\tilde{C}_i\frac{M_i}{2\pi i\hbar t}\left|\frac{\omega_i t}{Sin(\omega_i t)}\right|\lim_{N\to\infty}J^2N\left(\frac{t}{\pi}\right)^{2N}\prod_{n=1}^{N}1/n^2=\tilde{C}_iF_i^2(t), \qquad (i=1,2), \quad (C6)$$

where $\omega_i^2=\omega_{0i}^2-\lambda/M_i$.

In Eq.(C6) we used the property

$$\lim_{N\to\infty}J\sqrt{N}\left(\frac{t}{\pi}\right)^N\prod_{n=1}^{N}1/n=1, \quad (C7)$$

see, for example [46].

Thus, we obtain the expression for the fluctuational integral

$$G = G_1 G_2 = \frac{M_1\omega_1}{2\pi i\hbar\,|\,Sin(\omega_1 t)\,|}\frac{M_2\omega_2}{2\pi i\hbar\,|\,Sin(\omega_2 t)\,|} = \tilde{C}_1\tilde{C}_2 F_1^2(t)F_2^2(t)\,, \quad \text{(C8)}$$

where $F(t)$ is the wave function amplitude for the undamped case [45, 46] but with renormalized eigenfrequencies of two oscillators due to their coupling.

## Appendix D. Time dependent functions in Eq.(56).

Here we repersent the functions with the terms only linear with respect the coupling constant $\lambda$

$$\begin{aligned} f_1(t) &= D_9 + D_9' + \Pi_6 - \frac{D_3 D_4 (D_{11} + D_{11}' + \Pi_{14})}{D_4^2 + 4a_1\hbar(C_1 + \hbar a_1)}, \\ f_2(t) &= D_{10} + D_{10}' + \Pi_7 - \frac{D_3' D_4' (D_{12} + D_{12}' + \Pi_{15}) + 2\hbar a_2 E_1 D_3'}{D_4'^2 + 4a_2\hbar(C_2 + \hbar a_2)}. \end{aligned} \quad \text{(D1)}$$

where we omitted the argument $t$ of functions $C(t), D(t), E(t)$ for brevity.

$$\begin{aligned} E_1(t) &= \frac{2M_1\gamma_1}{\pi} r_2 \int_0^\infty d\omega\,\omega\, Coth\left(\frac{\hbar\omega}{2k_B T_1}\right) \\ &\int_0^t\int_0^\tau ds d\tau\{-2[m_1^2 \sin(\Omega_1\tau)\cos[\omega(\tau-s)]\sin(\Omega_1 s) + \cos(\Omega_1\tau)\cos[\omega(\tau-s)]\cos(\Omega_1 s)]e^{\delta_1(\tau+s)} + \\ &+2m_1[\cos(\Omega_1\tau)\cos[\omega(\tau-s)]\sin(\Omega_1 s) + \sin(\Omega_1\tau)\cos[\omega(\tau-s)]\cos(\Omega_1 s)]e^{\delta_1(\tau+s)} + g_1(\tau,s)\} + \\ &+\frac{2M_2\gamma_2}{\pi} r_1 \int_0^\infty d\omega\,\omega\, Coth\left(\frac{\hbar\omega}{2k_B T_2}\right) \\ &\int_0^t\int_0^\tau ds d\tau\{-2[m_2^2 \sin(\Omega_2\tau)\cos[\omega(\tau-s)]\sin(\Omega_2 s) + \cos(\Omega_2\tau)\cos[\omega(\tau-s)]\cos(\Omega_2 s)]e^{\delta_2(\tau+s)} + \\ &+2m_2[\cos(\Omega_2\tau)\cos[\omega(\tau-s)]\sin(\Omega_2 s) + \sin(\Omega_2\tau)\cos[\omega(\tau-s)]\cos(\Omega_2 s)]e^{\delta_2(\tau+s)} + g_1(\tau,s)\} \end{aligned}, \quad \text{(D2)}$$

where

$$\begin{aligned} g_1(\tau,s) &= m_1 m_2[\sin(\Omega_2\tau)\cos[\omega(\tau-s)]\sin(\Omega_1 s)e^{\delta_2\tau+\delta_1 s} + \sin(\Omega_1\tau)\cos[\omega(\tau-s)]\sin(\Omega_2 s)e^{\delta_1\tau+\delta_2 s}] - \\ &-m_2[\sin(\Omega_2\tau)\cos[\omega(\tau-s)]\cos(\Omega_1 s)e^{\delta_2\tau+\delta_1 s} + \cos(\Omega_1\tau)\cos[\omega(\tau-s)]\sin(\Omega_2 s)e^{\delta_1\tau+\delta_2 s}] + \\ &+[\cos(\Omega_2\tau)\cos[\omega(\tau-s)]\cos(\Omega_1 s)e^{\delta_2\tau+\delta_1 s} + \cos(\Omega_1\tau)\cos[\omega(\tau-s)]\cos(\Omega_2 s)e^{\delta_1\tau+\delta_2 s}] - \\ &-m_1[\cos(\Omega_2\tau)\cos[\omega(\tau-s)]\sin(\Omega_1 s)e^{\delta_2\tau+\delta_1 s} + \sin(\Omega_1\tau)\cos[\omega(\tau-s)]\cos(\Omega_2 s)e^{\delta_1\tau+\delta_2 s}]. \end{aligned} \quad \text{,(D3)}$$

$$E_2(t)=\frac{2M_1\gamma_1}{\pi}r_2\int_0^\infty d\omega\,\omega Coth\left(\frac{\hbar\omega}{2k_BT_1}\right)\int_0^t\int_0^\tau dsd\tau\{2\bar{n}_1m_1\sin(\Omega_1\tau)\cos[\omega(\tau-s)]\sin(\Omega_1 s)e^{\delta_1(\tau+s)}$$
$$-\bar{n}_1[\cos(\Omega_1\tau)\cos[\omega(\tau-s)]\sin(\Omega_1 s)+\sin(\Omega_1\tau)\cos[\omega(\tau-s)]\cos(\Omega_1 s)]e^{\delta_1(\tau+s)}+g_2(\tau,s)\}+$$
$$+\frac{2M_2\gamma_2}{\pi}r_1\int_0^\infty d\omega\,\omega Coth\left(\frac{\hbar\omega}{2k_BT_2}\right)\int_0^t\int_0^\tau dsd\tau\{2\bar{n}_2m_2\sin(\Omega_2\tau)\cos[\omega(\tau-s)]\sin(\Omega_2 s)e^{\delta_2(\tau+s)}$$
$$-\bar{n}_2[\cos(\Omega_2\tau)\cos[\omega(\tau-s)]\sin(\Omega_2 s)+\sin(\Omega_2\tau)\cos[\omega(\tau-s)]\cos(\Omega_2 s)]e^{\delta_2(\tau+s)}+g_2(\tau,s)\} \quad \text{, (D4)}$$

where

$$g_2(\tau,s)=-\bar{n}_2m_1[\sin(\Omega_2\tau)\cos[\omega(\tau-s)]\sin(\Omega_1 s)e^{\delta_2\tau+\delta_1 s}+\sin(\Omega_1\tau)\cos[\omega(\tau-s)]\sin(\Omega_2 s)e^{\delta_1\tau+\delta_2 s}]$$
$$+\bar{n}_2[\sin(\Omega_2\tau)\cos[\omega(\tau-s)]\cos(\Omega_1 s)e^{\delta_2\tau+\delta_1 s}+\cos(\Omega_1\tau)\cos[\omega(\tau-s)]\sin(\Omega_2 s)e^{\delta_1\tau+\delta_2 s}]. \quad \text{(D5)}$$

The function $E_3(t)$ has similar structure as $E_2(t)$ but corresponding function $g_3(\tau,s)$ reads

$$g_3(\tau,s)=-\bar{n}_1m_2[\sin(\Omega_2\tau)\cos[\omega(\tau-s)]\sin(\Omega_1 s)e^{\delta_2\tau+\delta_1 s}+\sin(\Omega_1\tau)\cos[\omega(\tau-s)]\sin(\Omega_2 s)e^{\delta_1\tau+\delta_2 s}]$$
$$+\bar{n}_1[\cos(\Omega_2\tau)\cos[\omega(\tau-s)]\sin(\Omega_1 s)e^{\delta_2\tau+\delta_1 s}+\sin(\Omega_1\tau)\cos[\omega(\tau-s)]\cos(\Omega_2 s)e^{\delta_1\tau+\delta_2 s}], \quad \text{(D6)}$$

$$E_4(t)=\frac{2M_1\gamma_1}{\pi}r_2\int_0^\infty d\omega\,\omega Coth\left(\frac{\hbar\omega}{2k_BT_1}\right)\int_0^t\int_0^\tau dsd\tau\{-2\bar{n}_1^2\sin(\Omega_1\tau)\cos[\omega(\tau-s)]\sin(\Omega_1 s)e^{\delta_1(\tau+s)}$$
$$+\bar{n}_1\bar{n}_2[\sin(\Omega_2\tau)\cos[\omega(\tau-s)]\sin(\Omega_1 s)e^{\delta_2\tau+\delta_1 s}+\sin(\Omega_1\tau)\cos[\omega(\tau-s)]\sin(\Omega_2 s)e^{\delta_1\tau+\delta_2 s}]\}+$$
$$+\frac{2M_2\gamma_2}{\pi}r_1\int_0^\infty d\omega\,\omega Coth\left(\frac{\hbar\omega}{2k_BT_2}\right)\int_0^t\int_0^\tau dsd\tau\{-2\bar{n}_2^2\sin(\Omega_2\tau)\cos[\omega(\tau-s)]\sin(\Omega_2 s)e^{\delta_2(\tau+s)}$$
$$+\bar{n}_1\bar{n}_2[\sin(\Omega_2\tau)\cos[\omega(\tau-s)]\sin(\Omega_1 s)e^{\delta_2\tau+\delta_1 s}+\sin(\Omega_1\tau)\cos[\omega(\tau-s)]\sin(\Omega_2 s)e^{\delta_1\tau+\delta_2 s}]\} \quad \text{. (D7)}$$

**Figure captions:**

**Figure 1.** A schematic sketch of the problem studied. Two selected harmonic oscillators and two independent reservoirs 1 and 2 of harmonic oscillators are coupled at times $t \geq 0$. The couplings are shown by thin solid lines.

**Figure 2.** Temporal dynamics of normalized variances $\sigma_{1,2}^2(t)/\sigma_{1,2}^2(FDT)$ for the first (curve 1) and second (curve 2) oscillators connected with different thermostats at equal temperatures $T_1 = T_2 = 300K$ in accordance with Eqs.(54), (55) and (60). In figure 2(a) the initial variances are $\sigma_{1,2}^2(t=0) = \hbar/2m_{1,2}\omega_{01,02}$ corresponding to the natural case of the cold system at $T_1 = T_2 = 0K$ in Eq.(63) and in figure 2(b) the initial variances $\sigma_1^2(t=0) = \hbar/2m_1\omega_{01}$ and $\sigma_2^2(t=0) = 10(\hbar/2m_2\omega_{02})$ corresponding to the arbitrary state of the second oscillator. Normalization was done to the variances of oscillators in equilibrium at $300K$ in accordance with the Eq.(60) and indicated as $\sigma_{1,2}^2(FDT)$.

**Figure 3.** Temporal behavior of normalized variances $\sigma_{1,2}^2(t)/\sigma_{1,2}^2(FDT)$ for the first 1 and second 2 oscillators connected with different thermostats at different temperatures $T_1 = 200K$ and $T_2 = 700K$ in accordance with Eqs.(54), (55) and (60) . In figure 3(a) the initial variances are $\sigma_{1,2}^2(t=0) = \hbar/2m_{1,2}\omega_{01,02}$ and in figure 3(b) the initial variances $\sigma_1^2(t=0) = \hbar/2m_1\omega_{01}$ and $\sigma_2^2(t=0) = 10(\hbar/2m_2\omega_{02})$ . Normalization was done to the variances of oscillators in equilibrium at $200K$ and $700K$, respectively.

**Figure 4.** Temporal dynamics of normalized variances $\sigma_{1,2}^2(t)/\sigma_{1,2}^2(FDT)$ for the first 1 and second 2 oscillators connected with different thermostats at absolute zero temperature $T_1 = T_2 = 0K$ in accordance with Eqs.(54), (55) and (60). In figure 4(a) the initial variances are $\sigma_{1,2}^2(t=0) = \hbar/2m_{1,2}\omega_{01,02}$ corresponding to the subsystems at $T_1 = T_2 = 0K$ and normalization was done to the variances of oscillators in equilibrium at $T_1 = T_2 = 0K$ in accordance with the Eq.(60). Normalization was also done to the variances at absolute zero temperature, with use of Eq.(60).

**Figure 5.** Normalized variances $\sigma_{1,2}^2(t)/\sigma_{1,2}^2(FDT)$ as the functions of time in accordance with Eqs.(54), (55) and (60) for the first in figure 5(a) and for the second in figure 5(b) oscillators connected with different thermostats at different temperatures $T = 10K$ , $T = 100K$ , $T = 300K$ and $T = 1000K$ (from the upper to low curve in each figures). Parameters of oscillators are $m_2 = 5m_1$, $\omega_{02} = 3\omega_{01}$ , $\gamma_{1,2} = 0.01\omega_{01,02}$, where $m_1 = 10^{-23}g$ , $\omega_{01} = 10^{13} rad/s$ . Initial variances are $\sigma_{1,2}^2(t=0) = \hbar/2m_{1,2}\omega_{01,02}$ and normalizations were done to the variances in equilibrium at corresponding temperatures.

**Figure 6.** Temporal dynamics of the normalized covariance $\beta_{12}(t)/\sqrt{\sigma_1^2\sigma_2^2(FDT)}$ for the coupled oscillators connected with different thermostats at different temperatures $T_1 = 200K$ and $T_2 = 700K$ in accordance with Eqs.(56),(60). Characteristics of two coupled oscillators are the same for each figure and indicated in the text. Coupling parameter $\rho$ is shown in inserts for each picture. Normalization was done to the variances of oscillators in equilibrium at $200K$ and $700K$ respectively, with use of Eq.(60).